\documentclass[amsmath,amssymb,aps,prd,reprint,superscriptaddress]{revtex4-1}

\usepackage{graphicx}
\usepackage{dcolumn}
\usepackage{bm}

\begin{document}


\title{Control of a filter cavity with coherent control sidebands}


\author{Naoki Aritomi}
\email[]{aritomi@granite.phys.s.u-tokyo.ac.jp}
\affiliation{Department of Physics, University of Tokyo, 7-3-1 Hongo, Tokyo, 113-0033, Japan}
\affiliation{National Astronomical Observatory of Japan, 2-21-2 Osawa, Mitaka, Tokyo, 181-8588, Japan}

\author{Matteo Leonardi}
\affiliation{National Astronomical Observatory of Japan, 2-21-2 Osawa, Mitaka, Tokyo, 181-8588, Japan}

\author{Eleonora Capocasa}
\affiliation{National Astronomical Observatory of Japan, 2-21-2 Osawa, Mitaka, Tokyo, 181-8588, Japan}

\author{Yuhang Zhao}
\affiliation{National Astronomical Observatory of Japan, 2-21-2 Osawa, Mitaka, Tokyo, 181-8588, Japan}
\affiliation{Department of Astronomical Science, SOKENDAI, 2-21-2 Osawa, Mitaka, Tokyo, 181-8588, Japan}

\author{Raffaele Flaminio}
\affiliation{Laboratoire d'Annecy-le-Vieux de Physique des Particules (LAPP),\\Universit Savoie Mont Blanc, CNRS/IN2P3, F-74941 Annecy-le-Vieux, France}
\affiliation{National Astronomical Observatory of Japan, 2-21-2 Osawa, Mitaka, Tokyo, 181-8588, Japan}


\date{\today}

\begin{abstract}
For broadband quantum noise reduction of gravitational-wave detectors, frequency-dependent squeezed vacuum states realized using a filter cavity is a mature technique and will be implemented in Advanced LIGO and Advanced Virgo from the fourth observation run. To obtain the benefit of frequency-dependent squeezing, detuning and alignment of the filter cavity with respect to squeezed vacuum states must be controlled accurately. To this purpose, we suggest a new length and alignment control scheme, using  coherent control sidebands which are already used to control the squeezing angle. Since both squeezed vacuum states and coherent control sidebands have the same mode matching conditions and almost the same frequency, detuning and alignment of the filter cavity can be controlled accurately with this scheme. In this paper, we show the principle of this scheme and  its application to a gravitational-wave detector.
\end{abstract}


\maketitle

\section{Introduction\label{introduction}}
Gravitational waves (GW) were first detected by Advanced LIGO in 2015 \cite{PhysRevLett.116.061102} and since then many more GW observations have been performed by Advanced LIGO and Advanced Virgo \cite{PhysRevX.9.031040}. To increase the number of detections, the sensitivity of the detectors must be constantly improved. One of the main noise sources which limits the sensitivity of GW detectors is the so-called quantum noise. Quantum noise is divided into shot noise, which limits the sensitivity at high frequency and radiation pressure noise, which limits the sensitivity at low frequency. An effective way to reduce quantum noise is to inject squeezed vacuum states into the interferometer \cite{PhysRevD.23.1693}. The reduction of quantum noise with squeezing was realized for the first time at GEO600 \cite{NatPhys.7.12.962}, and it has been recently implemented also in Advanced LIGO and Advanced Virgo since the beginning of the third observation run (O3) \cite{PhysRevLett.123.231107, PhysRevLett.123.231108}. However, conventional frequency-independent phase squeezed vacuum states increases radiation pressure noise at low frequency while it reduces shot noise at high frequency. For broadband quantum noise reduction, frequency-dependent squeezing produced with a filter cavity is the most promising technique \cite{PhysRevD.65.022002}. Advanced LIGO and Advanced Virgo plan to implement frequency-dependent squeezing with 300 m filter cavities from the fourth observation run (O4) \cite{arXiv:1304.0670}. In order to achieve the frequency dependence below 100 Hz, the cavity has to be operated in a detuned configuration which means off resonance of the carrier and it needs a storage time of about 3 ms.
\par Demonstration of frequency-dependent squeezing below 100 Hz, necessary for broadband quantum noise reduction in GW detectors, has been recently achieved \cite{PhysRevLett.124.171101, PhysRevLett.124.171102}. 
\par One of main challenges in the production of frequency-dependent squeezing by using filter cavities is the length and alignment control of the filter cavity itself. In fact, since squeezing is a vacuum state with no coherent amplitude, it is not suitable to provide the error signals necessary for the control. The use of auxiliary fields is therefore needed.
 In previous experiments \cite{PhysRevLett.124.171101}, the filter cavity was controlled with an auxiliary green field with a wavelength of 532 nm while the squeezed field is at the GW detector laser wavelength, 1064 nm. However, controlling length and alignment of the filter cavity with the green field does not ensure the alignment of squeezed field to the filter cavity, since the overlap of the green and squeezed field can drift. In addition, fluctuation of the relative phase delay between green and infrared field induced by anisotropies or temperature dependency of the cavity mirror coating can lead to a detuning fluctuation \cite{coating_book}. Another challenge of the filter cavity control with the green field is that phase/frequency noise on the green field creates real length noise in the filter cavity due to feedback control \cite{PhysRevLett.124.171102}.
\par The squeezed field is produced by a parametric down-conversion process inside an optical parametric oscillator (OPO). The use of an auxiliary field, which resonates inside the OPO, ensures that it is perfectly spatially overlapped with the squeezed field.
 For the length control of the filter cavity, a recent work has successfully tested a scheme which uses an additional auxiliary field injected into the OPO with a small frequency offset with respect to the squeezed field \cite{PhysRevLett.124.171102}. 
\par In this paper, we suggest a new length and alignment control scheme whose error signal is provided by the so-called coherent control (CC) field. Such field is included in all the squeezed vacuum sources for GW detectors and it is used to control the squeezing angle \cite{PhysRevA.75.043814}. Since the coherent control sidebands (CCSB) are produced inside the OPO together with the squeezed vacuum states, they have the same mode matching conditions and almost the same frequency. The relative frequency of carrier and CCSB can be controlled accurately with a frequency offset phase locked loop and can be tuned so that carrier is properly detuned. Such difference is only a few MHz which makes any possible effect due to the coating negligible. 
Therefore, length and alignment control with CCSB ensure proper detuning and alignment of the squeezed vacuum states to the filter cavity.
\par This paper is organized as follows: in section \ref{A} and \ref{B}, the error signal for the length and alignment control of the filter cavity are theoretically derived. In section \ref{C}, the application of such control scheme to a GW detector is presented. In section \ref{D}, the coupling between the coherent control loop and the filter cavity length control loop is studied. In section \ref{E}, reshaping of frequency-dependent phase noise and an updated squeezing degradation budget with this control scheme are presented. In section \ref{III},  the computation of noise requirements to ensure the feasibility of such technique is presented.

\section{Principle}
\subsection{Filter cavity length signal}
\label{A}
When coherent control field, which is detuned by $\Omega_\mathrm{cc}$ with respect to carrier frequency $\omega_0$, is injected into OPO, a sideband which is detuned by $-\Omega_\mathrm{cc}$ is generated by the pump field whose frequency is 2$\omega_0$ \cite{PhysRevA.75.043814}. 
The coherent control field passing through OPO can be written as \cite{Eric2016}
\begin{eqnarray}
E_\mathrm{cc} &=& a_\mathrm{cc} \dfrac{1}{(1-x^2)} e^{i(\omega_0+\Omega_\mathrm{cc})t + i\phi_\mathrm{CC}} \nonumber\\
&+& a_\mathrm{cc} \dfrac{x}{(1-x^2)} e^{i(\omega_0-\Omega_\mathrm{cc})t + i(\phi_\mathrm{CC}+2 \phi_\mathrm{pump})} \label{generated cc}
\end{eqnarray}
where $a_\mathrm{cc}$ is the amplitude of the coherent control field without the pump field, $x$ is the OPO nonlinear factor and $\phi_\mathrm{CC}$, $\phi_\mathrm{pump}$ are the common and relative phase of the coherent control field and its sideband generated by OPO respectively. Note that (\ref{generated cc}) assumes that $\Omega_\mathrm{cc}$ is much lower than the OPO bandwidth $\gamma_\mathrm{OPO}$, $\Omega_\mathrm{cc} \ll \gamma_\mathrm{OPO}$. The OPO nonlinear factor $x$ can be written as 
\begin{eqnarray}
x = \sqrt{\dfrac{P_\mathrm{pump}}{P_\mathrm{th}}} = 1 - \frac{1}{\sqrt{g}}
\end{eqnarray}
where $P_\mathrm{pump}$ is the power of the pump field, $P_\mathrm{th}$ is the OPO threshold power, and $g$ is the nonlinear gain. $g$ and $\phi_\mathrm{pump}$ are determined by the amplitude and phase of the pump field, respectively. The nonlinear gain $g$ is related to the generated squeezing $\sigma_\mathrm{dB}$ without losses as follows \cite{Eric2016}:
\begin{eqnarray}
\sigma_\mathrm{dB} = 20\log_{10}(2\sqrt{g}-1)
\end{eqnarray}
$\phi_\mathrm{CC}$ and $\phi_\mathrm{pump}$ are controlled by the coherent control loops to control the squeezing angle. 
The squeezing angle is the relative phase between the local oscillator and the average of CCSB and can be written as
\begin{eqnarray}
\phi_\mathrm{sqz} = \phi_\mathrm{LO} - \phi_\mathrm{CC} -\phi_\mathrm{pump} \label{squeezing angle}
\end{eqnarray}
where $\phi_\mathrm{LO}$ is the phase of the local oscillator (LO). There are two coherent control loops (we call them CC1 and CC2 in this paper) and $\phi_\mathrm{pump}$ is kept constant by CC1 and the relative phase between LO and CC $\phi_\mathrm{LO} - \phi_\mathrm{CC}$ is kept constant by CC2 to make the squeezing angle $\phi_\mathrm{sqz}$ constant. In this paper, we assume that $\phi_\mathrm{pump}$ is kept 0, but has residual noise around 0, $\phi_\mathrm{pump} = \delta \phi_\mathrm{pump} \ll 1$.\par
\begin{figure}[t]
\begin{center}
\includegraphics[width=15cm,bb= 00 150 1370 550]{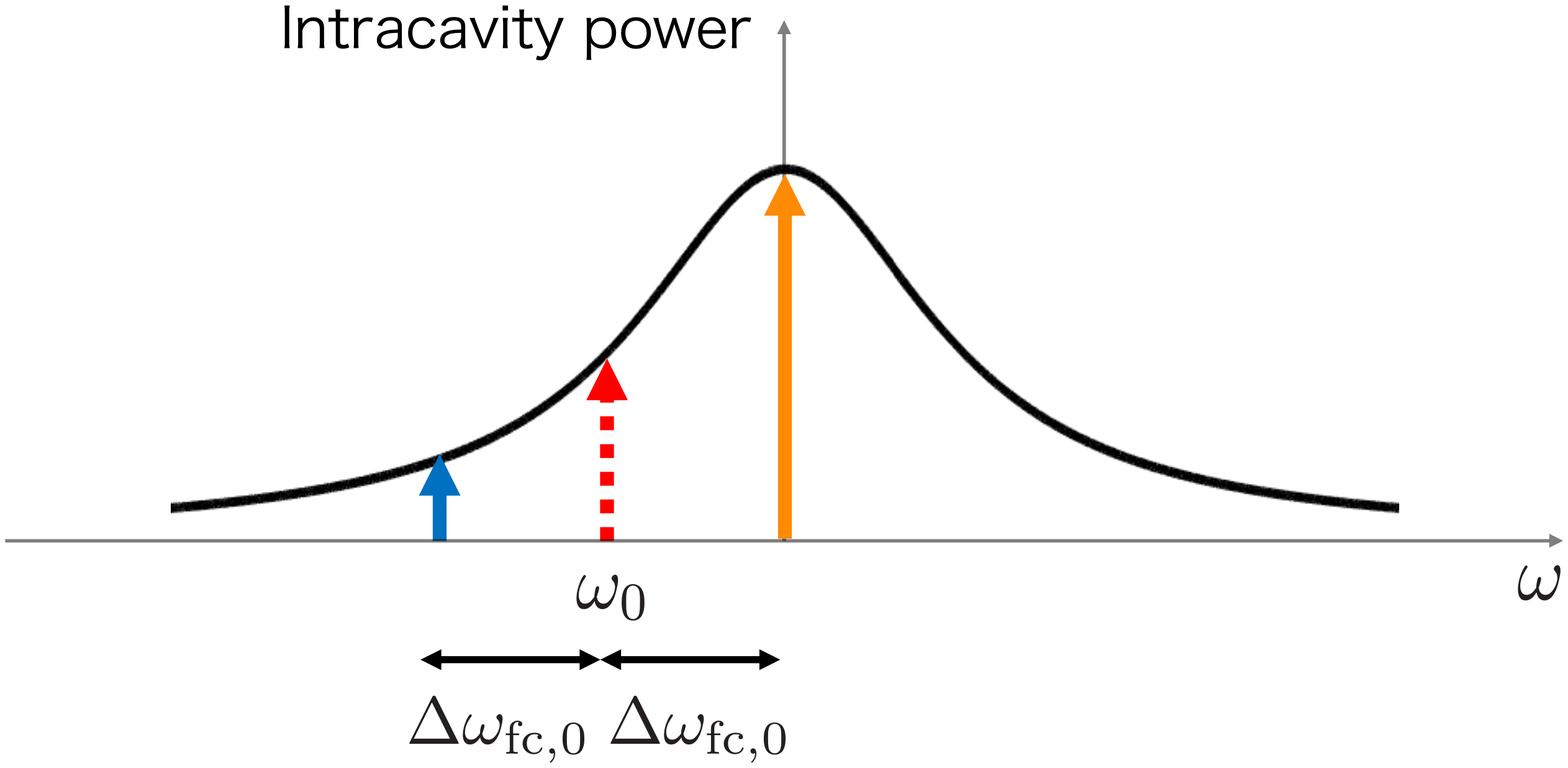}
\caption{Frequency relationship inside the filter cavity. Red dashed line is carrier, orange and blue lines are coherent control sidebands.}
\label{CC frequency}
\end{center}
\end{figure}
\begin{figure}[t]
\begin{center}
\includegraphics[width=12cm,bb= 00 10 1670 700]{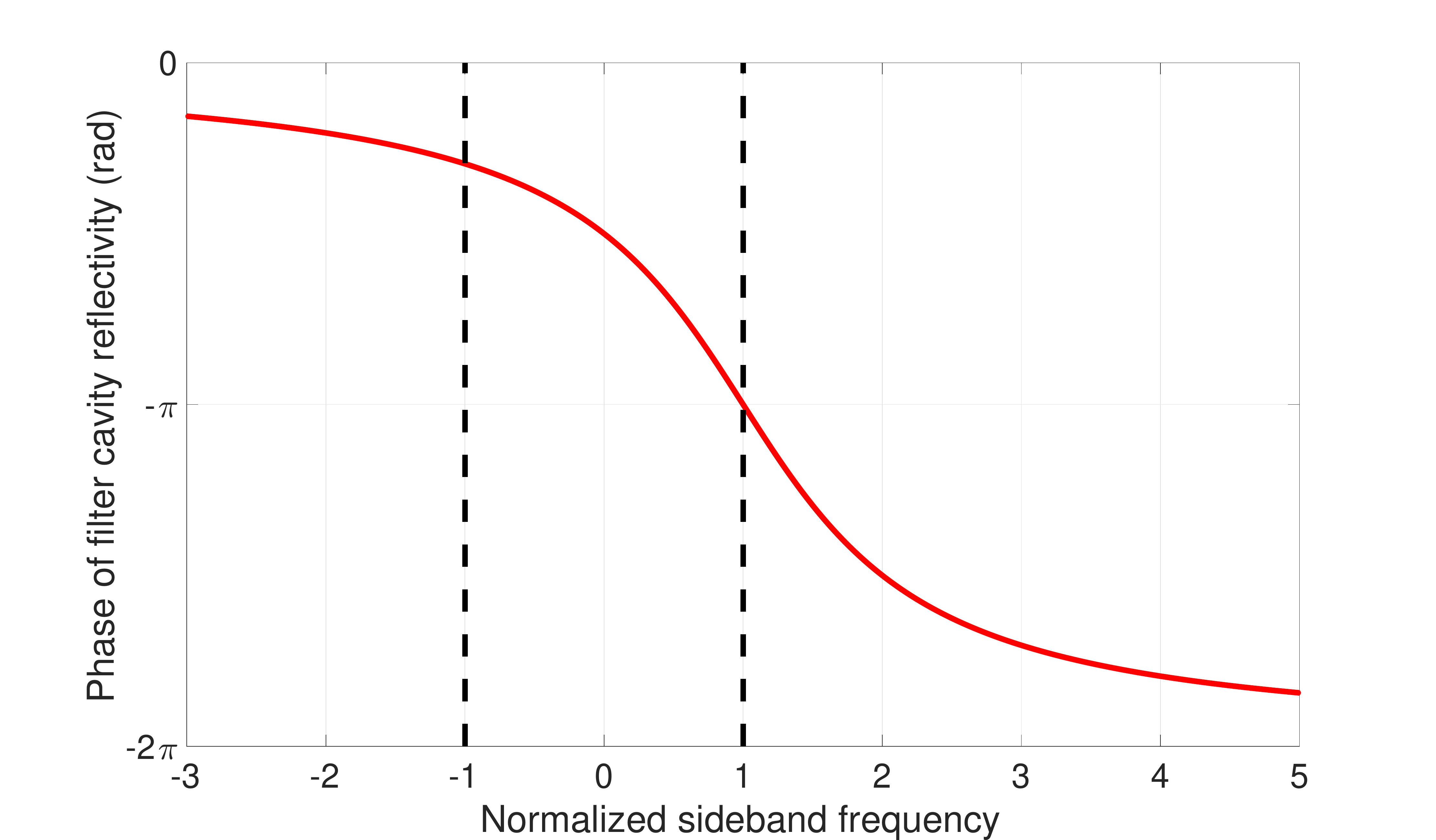}
\caption{Phase of the filter cavity reflectivity. The horizontal axis is the sideband frequency normalized with respect to $\Delta\omega_{\mathrm{fc},0}$. Black dashed lines are the sideband frequency of CCSB.}
\label{reflection}
\end{center}
\end{figure}

To obtain frequency-dependent squeezing from a filter cavity, the resonance of the filter cavity must be detuned properly from the carrier. By choosing the frequency of coherent control field ($\Omega_\mathrm{cc}$) as follows, the coherent control field can be resonant inside the filter cavity while the resonance of the filter cavity is properly detuned from the carrier (FIG. \ref{CC frequency}):
\begin{equation}
\Omega_\mathrm{cc} = n\times \omega_\mathrm{FSR} + \Delta \omega_{\mathrm{fc},0}
\end{equation}
where $n$ is an integer number, $\omega_\mathrm{FSR} = 2\pi f_\mathrm{FSR}=\pi c/L_\mathrm{fc}$ is the free spectral range of the filter cavity, $L_\mathrm{fc}$ is the filter cavity length and $\Delta \omega_{\mathrm{fc},0}$ is the optimal filter cavity detuning with respect to carrier. In this condition, the coherent control sideband at $-\Omega_\mathrm{cc}$ is detuned by $-2\Delta \omega_{\mathrm{fc},0}$ with respect to the filter cavity resonance and so almost reflected by the filter cavity. The phase of the filter cavity reflectivity is shown in FIG. \ref{reflection}. Since the phase of filter cavity reflectivity for the coherent control field which is on resonance is sensitive to the detuning fluctuation compared with the other sideband which is off resonance, the filter cavity length signal can be obtained by detecting the beat note of CCSB.

The coherent control sidebands reflected by the filter cavity is
\begin{eqnarray}
E_\mathrm{cc} &=& a_+ r_+ e^{i(\omega_0+\Omega_\mathrm{cc})t+i\phi_\mathrm{CC}} \nonumber\\
&+& a_- r_- e^{i(\omega_0-\Omega_\mathrm{cc})t + i(\phi_\mathrm{CC}+2\delta \phi_\mathrm{pump})}
\end{eqnarray}
where $a_\pm$ is
\begin{equation}
a_+ = a_\mathrm{cc} \dfrac{1}{(1-x^2)},\ a_- = a_\mathrm{cc} \dfrac{x}{(1-x^2)}
\end{equation}
and $r_\pm (\Delta \omega_\mathrm{fc}) = r_\mathrm{fc}(\pm \Delta \omega_{\mathrm{fc},0}, \Delta\omega_\mathrm{fc})$ is the complex reflectivity of the filter cavity and can be written as \cite{PhysRevD.90.062006}
\begin{eqnarray}
r_\mathrm{fc}(\pm \Delta \omega_{\mathrm{fc},0}, \Delta\omega_\mathrm{fc})  \simeq 1 - \dfrac{2-\epsilon}{1+i\xi(\pm \Delta \omega_{\mathrm{fc},0},\Delta\omega_\mathrm{fc})} \label{r_fc}
\end{eqnarray}
where
\begin{eqnarray}
\epsilon &=& \dfrac{f_\mathrm{FSR}}{\gamma_\mathrm{fc}}\Lambda^2_\mathrm{rt}\\
\xi (\pm \Delta \omega_{\mathrm{fc},0},\Delta\omega_\mathrm{fc}) &=& \dfrac{\pm \Delta \omega_{\mathrm{fc},0} 
-\Delta\omega_\mathrm{fc}}{\gamma_\mathrm{fc}}
\end{eqnarray}
with $\gamma_\mathrm{fc}$ the filter cavity half-bandwidth and $\Lambda_\mathrm{rt}^2$ the filter cavity round trip losses. $\Delta\omega_\mathrm{fc}$ is a variable which represents the actual filter cavity detuning. Note that the approximation in (\ref{r_fc}) holds for a high finesse cavity near the resonance, $(\Omega-\Delta \omega_{\mathrm{fc},0})/f_\mathrm{FSR} \ll 1$ and $t_\mathrm{in}^2+\Lambda_\mathrm{rt}^2 \ll 1$ where $\Omega$ is the sideband frequency and $t_\mathrm{in}^2$ is the filter cavity input transmissivity.\par 
Amplitude and phase of the filter cavity reflectivity for CCSB can be written as
\begin{eqnarray}
\rho_{\pm}(\Delta\omega_\mathrm{fc})  &=& |r_\mathrm{fc}(\pm \Delta \omega_{\mathrm{fc},0},\Delta\omega_\mathrm{fc})| \nonumber\\
&=& \sqrt{1-\dfrac{(2-\epsilon)\epsilon}{1+\xi^2(\pm \Delta \omega_{\mathrm{fc},0},\Delta\omega_\mathrm{fc})}} \\
\alpha_{\pm}(\Delta\omega_\mathrm{fc}) &=& \mathrm{arg}\{r_\mathrm{fc}(\pm \Delta \omega_{\mathrm{fc},0},\Delta\omega_\mathrm{fc})\}\nonumber\\
&=& \mathrm{arg}\{-1+\epsilon+\xi^2 (\pm \Delta \omega_{\mathrm{fc},0},\Delta\omega_\mathrm{fc}) \nonumber\\
&& +\ i(2-\epsilon)\xi(\pm \Delta \omega_{\mathrm{fc},0},\Delta\omega_\mathrm{fc})\} \label{FC phase}
\end{eqnarray}

\begin{figure}[t]
\begin{center}
\includegraphics[width=12cm,bb= 00 10 1670 700]{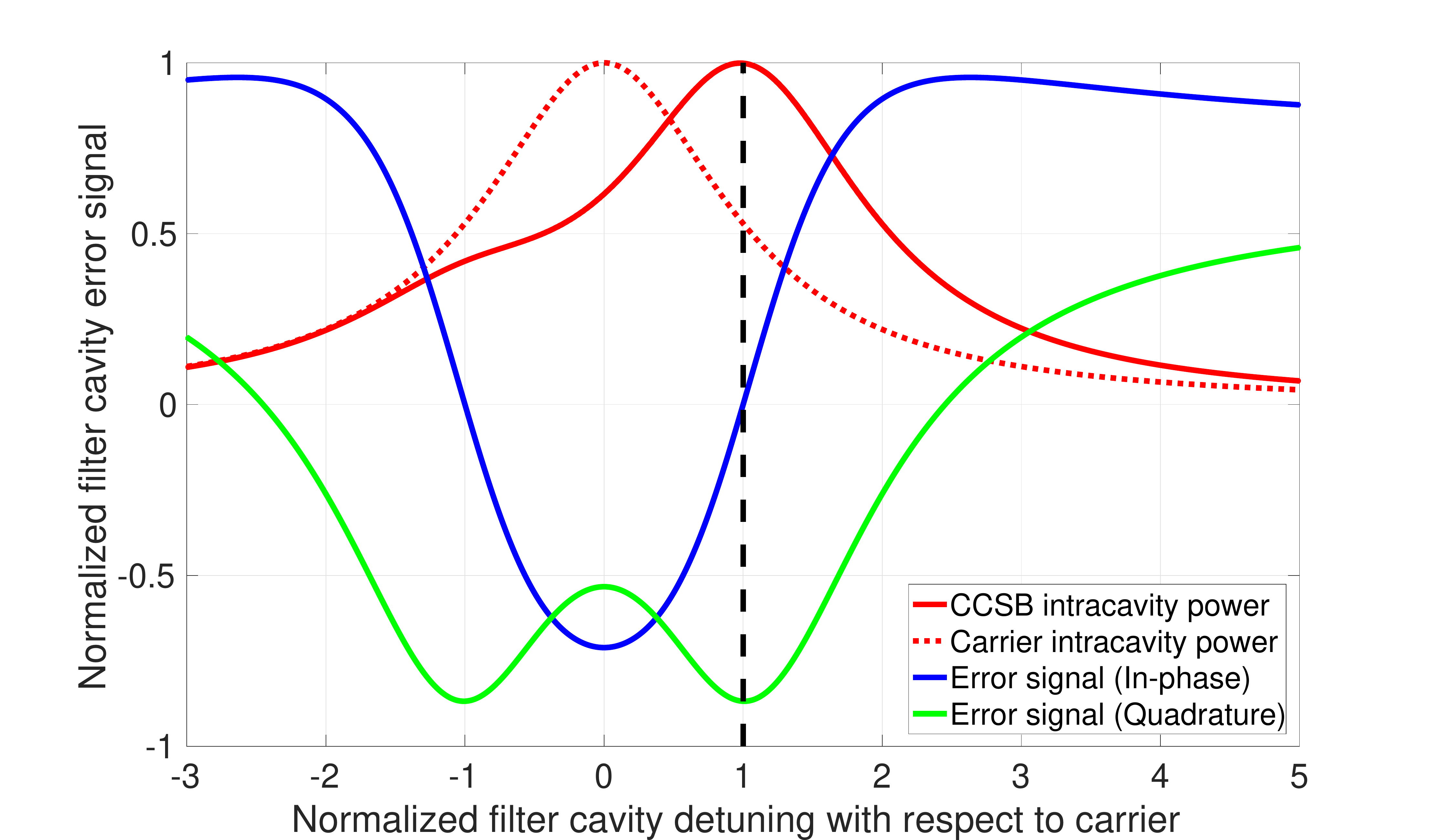}
\caption{Filter cavity length signal normalized with respect to $a_+ a_-$. The horizontal axis is the filter cavity detuning normalized with respect to $\Delta \omega_{\mathrm{fc},0}$. Red solid and dashed lines are intracavity power of CCSB and carrier in the filter cavity normalized with respect to their maximum intracavity power, respectively. Blue and green lines are the filter cavity length signal (In-phase and Quadrature). The filter cavity length signal (In-phase) becomes 0 when $\Delta\omega_\mathrm{fc} = \Delta\omega_{\mathrm{fc},0}$.}
\label{FC error}
\end{center}
\end{figure}

Filter cavity length signal can be obtained by detecting the beat note of the CCSB:
\begin{eqnarray}
P_\mathrm{cc} &=& \left| a_+ r_+ e^{i(\omega_0 + \Omega_\mathrm{cc})t} + a_- r_- e^{i(\omega_0 - \Omega_\mathrm{cc})t + i2\delta\phi_\mathrm{pump}}\right|^2 \nonumber \\
&=& (\mathrm{DC\ term})\ +\ 2a_+ a_- \mathrm{Re}\{ r_+ r_-^\ast e^{i(2\Omega_\mathrm{cc}t-2\delta\phi_\mathrm{pump})}\} \nonumber\\ \label{error signal}
\end{eqnarray}
Demodulating this signal by $\sin{(2\Omega_\mathrm{cc} t -\alpha_{-}(\Delta \omega_{\mathrm{fc},0})})$ (In-phase) and $\cos{(2\Omega_\mathrm{cc} t -\alpha_{-}(\Delta \omega_{\mathrm{fc},0}))}$ (Quadrature) and low-passing it, filter cavity length signal as a function of the filter cavity detuning $\Delta\omega_\mathrm{fc}$ is
\begin{eqnarray}
P_I &=& - a_+ a_- \rho_+ (\Delta\omega_\mathrm{fc}) \rho_- (\Delta\omega_\mathrm{fc}) \nonumber\\
&\times& \sin{(\alpha_+ (\Delta\omega_\mathrm{fc}) - \alpha_- (\Delta\omega_\mathrm{fc}) 
+  \alpha_- (\Delta\omega_{\mathrm{fc},0}) -2\delta \phi_\mathrm{pump})}\nonumber \\ \label{I phase} \\  
P_Q &=&  a_+ a_- \rho_+ (\Delta\omega_\mathrm{fc}) \rho_- (\Delta\omega_\mathrm{fc})\nonumber \\
&\times& \cos{(\alpha_+ (\Delta\omega_\mathrm{fc}) - \alpha_- (\Delta\omega_\mathrm{fc}) 
+  \alpha_- (\Delta\omega_{\mathrm{fc},0}) -2\delta \phi_\mathrm{pump})}\nonumber \\ \label{Q phase} 
\end{eqnarray}
The relative phase noise of CCSB $\delta\phi_\mathrm{pump}$ is a noise source for the filter cavity length signal. When $\delta\phi_\mathrm{pump} = 0$, the filter cavity length signal (\ref{I phase}), (\ref{Q phase}) normalized with respect to $a_+ a_-$ is shown in FIG. \ref{FC error}. The parameters used in this calculation are shown in TABLE \ref{table1}.

\begin{table}[t]
\caption{\label{table1}%
Parameters for 300 m filter cavity \cite{PhysRevD.93.082004}.
}
\begin{ruledtabular}
\begin{tabular}{lcc}
\textrm{Parameter}&
\textrm{Symbol}&
\textrm{Value}\\
\colrule
Filter cavity length & $L_\mathrm{fc}$ & 300 m\\
Filter cavity half-bandwidth& $\gamma_\mathrm{fc}$ & $2\pi\times57.3$ Hz\\
Filter cavity detuning & $\Delta\omega_{\mathrm{fc},0}$ & $2\pi\times54$ Hz \\
Filter cavity finesse & $\mathcal{F}$ & 4360\\
Filter cavity input mirror transmissivity & $t_\mathrm{in}^2$ & 0.00136\\
Filter cavity round trip losses & $\Lambda^2_\mathrm{rt}$ & 80 ppm\\
Injection losses & $\Lambda^2_\mathrm{inj}$ & 5 $\%$\\
Readout losses & $\Lambda^2_\mathrm{ro}$ & 5 $\%$\\
Mode-mismatch losses & $\Lambda^2_\mathrm{mmFC}$ & 2 $\%$\\
(squeezer-filter cavity) &&\\
Mode-mismatch losses & $\Lambda^2_\mathrm{mmLO}$ & 5 $\%$\\
(squeezer-local oscillator) &&\\
Frequency-independent phase  &$\delta\phi$& 30 mrad\\
noise (rms) &&\\
Filter cavity length noise (rms)  &$\delta L_\mathrm{fc}$& 1 pm\\
Generated squeezing &$\sigma_\mathrm{dB}$& 9 dB\\
Nonlinear gain &$g$& 3.6\\
\end{tabular}
\end{ruledtabular}
\end{table}

Filter cavity length noise $\delta L_\mathrm{fc}$ causes detuning noise $\delta \Delta \omega_\mathrm{fc}$ as follows,
\begin{eqnarray}
\delta \Delta \omega_\mathrm{fc} = \dfrac{\omega_0}{L_\mathrm{fc}}\delta L_\mathrm{fc}\label{detuning noise}
\end{eqnarray}
When $\Delta\omega_\mathrm{fc} = \Delta\omega_{\mathrm{fc},0} + \delta \Delta \omega_\mathrm{fc}$, phase response of the filter cavity reflectivity to the detuning noise can be calculated from (\ref{FC phase}),
\begin{eqnarray}
\delta \alpha (\Omega) &=& \left. \dfrac{d \alpha(\Omega, \Delta \omega_\mathrm{fc})}{d \Delta \omega_\mathrm{fc}} \right|_{\Delta \omega_\mathrm{fc}=\Delta \omega_{\mathrm{fc},0}} \delta \Delta \omega_\mathrm{fc} \nonumber \\
&\simeq& \left( \dfrac{(\Omega - \Delta \omega_{\mathrm{fc},0})^2}{\gamma_\mathrm{fc}^2}+1\right)^{-1}\dfrac{8\mathcal{F}}{\lambda}\delta L_\mathrm{fc} \label{phase response}
\end{eqnarray}
where $\alpha(\Omega, \Delta \omega_\mathrm{fc}) = \mathrm{arg}\{r_\mathrm{fc}(\Omega,\Delta\omega_\mathrm{fc})\}$, $\lambda$ is wavelength of carrier, and $\mathcal{F}$ is filter cavity finesse. Here we assumed $\epsilon \ll 1$, which is true for parameters shown in TABLE \ref{table1}. The filter cavity length signal (In-phase) (\ref{I phase}) is 
\begin{eqnarray}
&& \frac{P_I}{a_+ a_- \rho_+ \rho_-} \nonumber\\
&=& \mathrm{sin}\left(\alpha_+ (\Delta\omega_{\mathrm{fc},0}+\delta\Delta\omega_\mathrm{fc}) 
- \alpha_+ (\Delta\omega_{\mathrm{fc},0}) \right. \nonumber\\
&& - \alpha_- (\Delta\omega_{\mathrm{fc},0}+\delta\Delta\omega_\mathrm{fc}) +  \left. \alpha_- (\Delta\omega_{\mathrm{fc},0})\right)\nonumber\\
&\simeq& \delta \alpha (\Delta \omega_{\mathrm{fc},0}) - \delta \alpha (-\Delta \omega_{\mathrm{fc},0})  \nonumber\\
&=& 26\ \mathrm{mrad}  \left(\dfrac{1064\ \mathrm{nm}}{\lambda}\right) \left(\dfrac{\mathcal{F}}{4360}\right)\left(\dfrac{\delta L_\mathrm{fc}}{1\ \mathrm{pm}}\right) \label{FC error rms} 
\end{eqnarray}
Since the relative phase noise of CCSB $\delta \phi_\mathrm{pump}$ can be stabilized by CC1 below 1.7 mrad \cite{PhysRevLett.117.110801}, the residual filter cavity length signal (\ref{FC error rms}) can be obtained with a good enough SNR. Phase noise of an RF source for the demodulation also becomes a noise source for the filter cavity length signal. Typical phase noise of an RF source for the demodulation is several tens of $\mu$rad and much smaller than (\ref{FC error rms}).

 \subsection{Filter cavity alignment signal}
 \label{B}
 Coherent control sidebands can be also used to control the alignment of the filter cavity by wavefront sensing \cite{Morrison:94}.
 \par Misalignment of the filter cavity axis with respect to input beam axis and misalignment of immediately reflected beam axis with respect to the filter cavity axis can be represented in terms of dimensionless coupling factors $\gamma, \gamma_r$ as follows:
 \begin{eqnarray}
 \gamma &=& \delta x/\mathrm{w}_0 + i\delta\theta/\theta_0 \label{gamma}\\
\gamma_r &=& \delta x^{\prime}/\mathrm{w}_0 + i\delta\theta^{\prime}/\theta_0 \label{gamma_r}
 \end{eqnarray}
where $\mathrm{w}_0$ is the beam radius at the waist position and $\theta_0 = \lambda/\pi \mathrm{w}_0$ is the beam divergence. $\delta x$ and $\delta x^\prime$ represent the transverse displacement in $x$ axis direction measured at the waist position of the filter cavity axis with respect to the input beam axis and the immediately reflected beam axis with respect to the filter cavity axis, respectively. $\delta \theta,\ \delta \theta^\prime$ represent the tilt around $y$ axis of the filter cavity axis with respect to the input beam axis and the immediately reflected beam axis with respect to the filter cavity axis, respectively. Here, $z$ axis is the beam axis and $y$ axis is orthogonal to the $x, z$ axis. $z=0$ is the beam waist position. $\gamma, \gamma_r$ can be written in terms of input, end mirror misalignment of the filter cavity as follows:
\begin{eqnarray}
\gamma &=& \dfrac{R}{2}\dfrac{\delta\theta_I + \delta\theta_E}{\mathrm{w}_0} + i \dfrac{R}{2R-L_\mathrm{fc}} \dfrac{\delta\theta_I - \delta\theta_E}{\theta_0} \label{gamma mirrors} \\
\gamma_r &=& \dfrac{ L_\mathrm{fc} \delta \theta_I -\frac{R}{2}(\delta\theta_I + \delta\theta_E)}{\mathrm{w}_0} + i \dfrac{2\delta\theta_I -\frac{R}{2R-L_\mathrm{fc}}(\delta\theta_I - \delta\theta_E)}{\theta_0} \nonumber \\
&=& \left(\frac{L_\mathrm{fc}}{\mathrm{w}_0} + i \frac{2}{\theta_0}\right)\delta \theta_I - \gamma \label{gamma_r mirrors}
\end{eqnarray}
where $R$ is the radius of curvature of the input and end mirrors, $\delta \theta_I$ and $\delta \theta_E$ are angular misalignment of input and end mirrors. The direction of the input and end mirror misalignment is defined so that the positive direction of the misalignment causes the displacement of filter cavity axis in positive direction of $x$ axis.
\par We only treat $x$-axis misalignment (Hermite-Gaussian [HG] 10 mode) and calculations of $y$-axis misalignment (HG 01 mode) are entirely analogous.\par
The wavefront sensing (WFS) signal can be written as (see Appendix \ref{Appendix A})
\begin{eqnarray}
W_\mathrm{diff} &=& 2a_+ a_- \sqrt{\frac{2}{\pi}}\mathrm{Re} (\{-r_+ r_-^* (e^{-i\eta} \gamma_r^* +e^{i\eta}\gamma_r) \nonumber\\
&&-  r_+ e^{-i\eta}\gamma - r_-^* e^{i\eta}\gamma^* \}e^{i2\Omega_\mathrm{cc} t})
\label{W QPD}
\end{eqnarray}
where $r_\pm = r_\pm (\Delta \omega_{\mathrm{fc},0})$ and $\eta$ is the gouy phase. Demodulating (\ref{W QPD}) by $\sin{(2\Omega_\mathrm{cc} t -\alpha_{-}(\Delta \omega_{\mathrm{fc},0})})$ (In-phase) and low-passing it, first term of (\ref{W QPD}), which is proportional to filter cavity length signal, will disappear. \par
\begin{figure}[t]
\begin{center}
\includegraphics[width=12cm,bb= 00 10 1670 700]{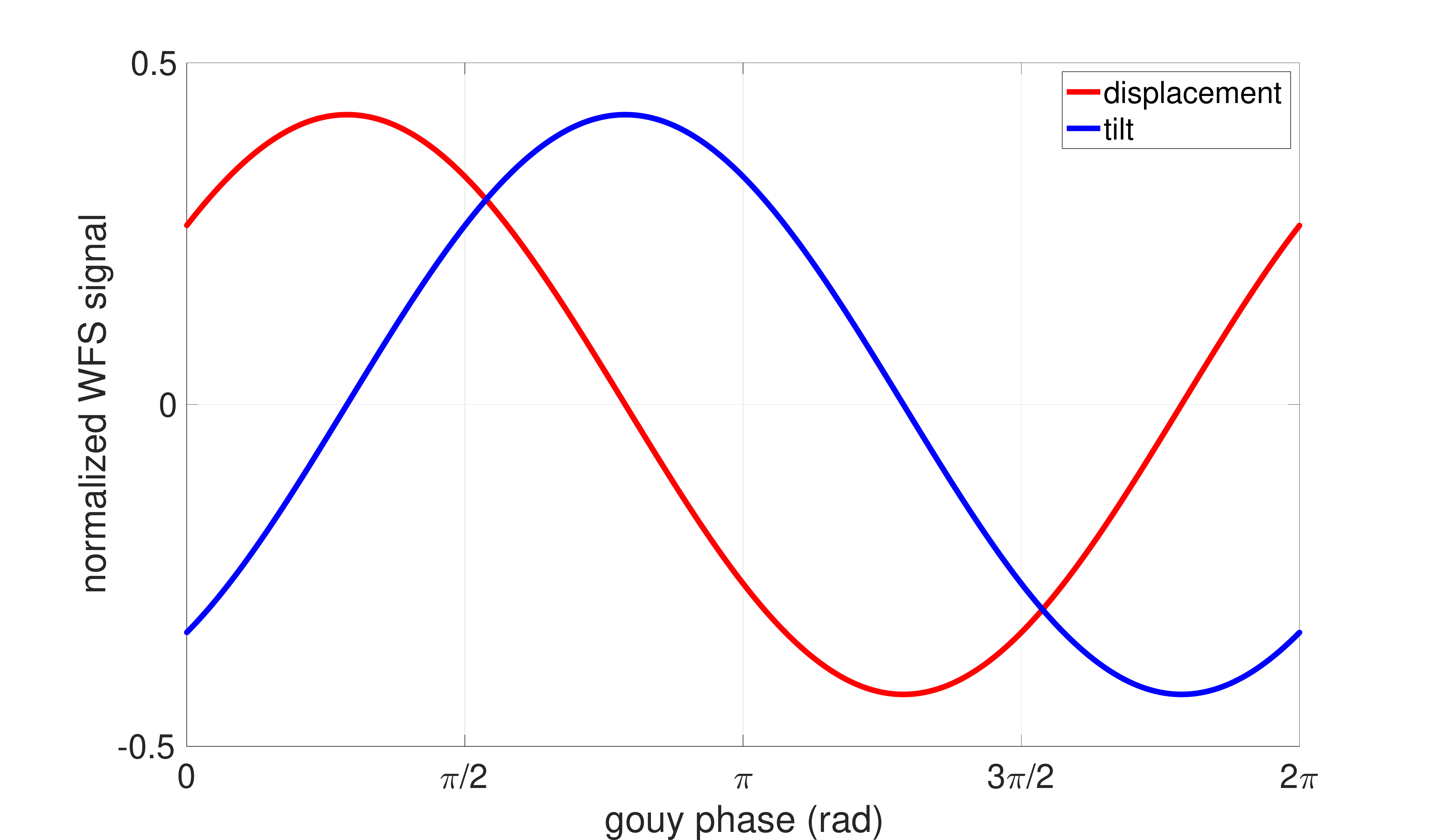}
\caption{Filter cavity WFS signal as a function of the gouy phase normalized with $\sqrt{\frac{2}{\pi}}a_+a_-$. Red line is displacement signal with $\delta x = \mathrm{w}_0$ and blue line is tilt signal with $\delta\theta = \theta_0$.}
\label{CC WFS}
\end{center}
\end{figure}

WFS signal after the demodulation is
\begin{eqnarray}
W_I &=& \sqrt{\frac{2}{\pi}}a_+a_- \nonumber\\
&\times& \{ \mathrm{Re} (r_+ e^{-i\eta}\gamma + r_-^* e^{i\eta}\gamma^* )\sin{\alpha_{-}(\Delta \omega_{\mathrm{fc},0})}\nonumber\\
&+&\ \mathrm{Im} (r_+ e^{-i\eta}\gamma + r_-^* e^{i\eta}\gamma^* )\cos{\alpha_{-}(\Delta \omega_{\mathrm{fc},0})} \} \label{WFS signal} 
\end{eqnarray}
WFS signal is shown in FIG. \ref{CC WFS}. Displacement and tilt signal of the filter cavity can be obtained with two different gouy phases.

The fourth term in (\ref{beat WFS}) is beat note of CCSB HG10 mode and becomes a noise source for the filter cavity length signal when misalignment of the filter cavity fluctuates. In order not to spoil the filter cavity length signal (\ref{FC error rms}), the requirement of maximum rms angular motion of filter cavity mirrors $\theta_\mathrm{max}$ should be $\theta_\mathrm{max}= 2.7$ $\mu$rad (see Appendix \ref{Appendix A}) and  this is achievable by double pendulum suspensions in KAGRA \cite{KOKEYAMA20181950}. By numerically calculating $|\gamma|^2$ as functions of input, end mirror misalignment, this requirement corresponds to $|\gamma|^2 =  0.01$. Since this requirement is more stringent than the requirement of mode-mismatch losses (squeezer-filter cavity) which is 2 $\%$, we set the requirement of $\gamma$ as  $|\gamma|^2 =  0.01$.
 
 \subsection{Experimental setup}
 \label{C}
\begin{figure}[h]
\begin{center}
\includegraphics[width=12cm,bb= 00 10 770 450]{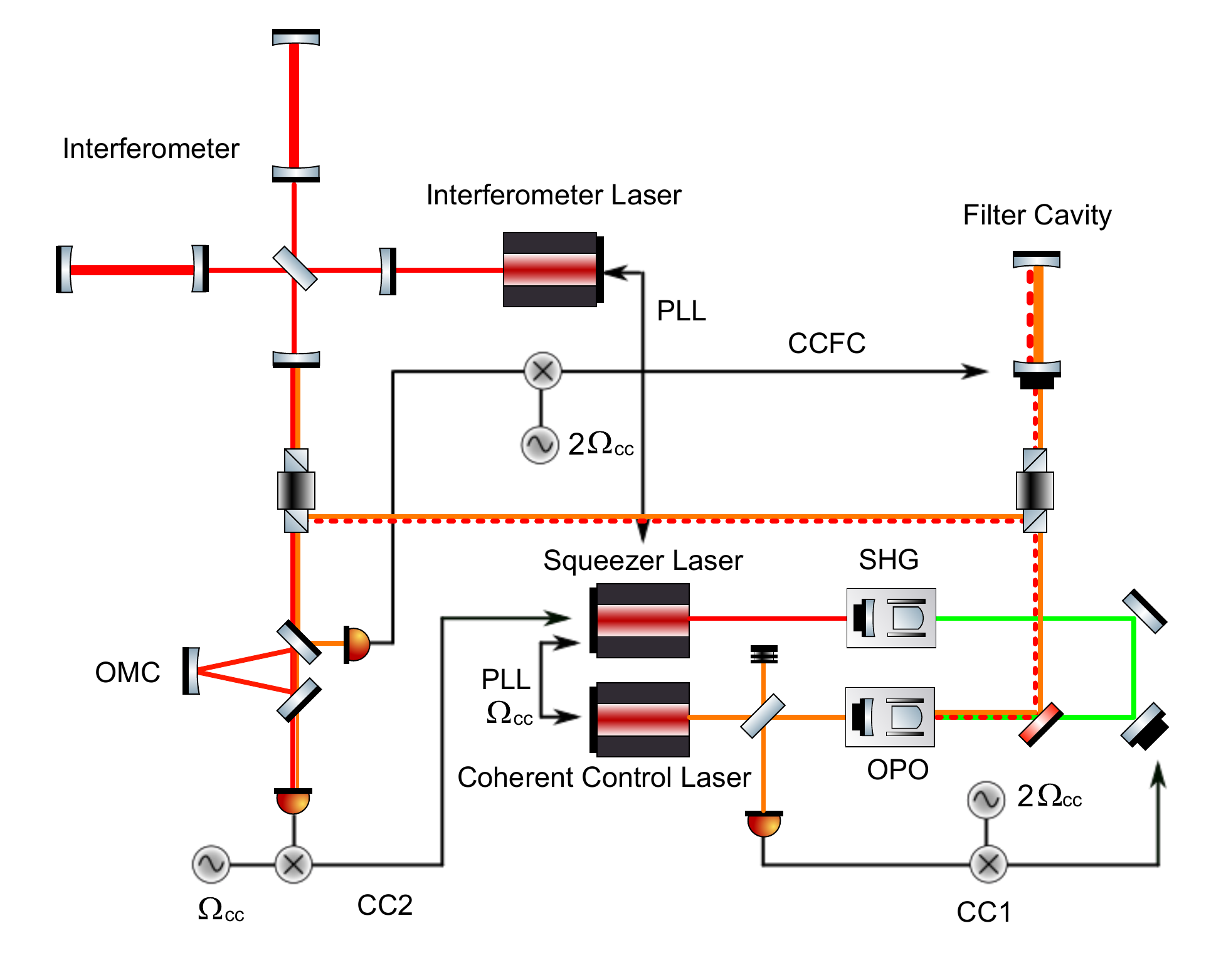}
\caption{An example of experimental setup. Red solid line is the carrier and red dashed line is the squeezed vacuum states. Orange line is the coherent control field. Green line is the green pump field which is produced by the second harmonic generator (SHG) and injected into the OPO.}
\label{CC setup}
\end{center}
\end{figure}

 An example of experimental setup when this scheme is implemented in GW detectors is shown in  FIG. \ref{CC setup}. There are three control loops which are CC1, CC2, and the filter cavity control loop with CCSB (we call it CCFC in this paper).
\par CCFC error signal can be obtained at output mode cleaner (OMC) reflection since CCSB are almost fully reflected by OMC while carrier almost transmits through OMC. The error signal is demodulated by 2$\Omega_\mathrm{cc}$ and fed back to the filter cavity length. The demodulation phase can be determined by injecting bright carrier field to the filter cavity and checking the carrier transmission and CCFC error signal at the same time as shown in FIG. \ref{FC error}. Fine tuning of the demodulation phase can be done by optimizing GW sensitivity. CC1 error signal to control the relative phase between the green field injected into OPO and the coherent control field can be obtained at OPO reflection and fed back to the green field path length. CC2 error signal to control the relative phase between the carrier and the coherent control field is obtained at OMC transmission and fed back to phase locked loop (PLL) between interferometer laser and squeezer laser \cite{Dooley:15}.

\subsection{Coherent control error signal}
\label{D}
Since either of CCSB enters the filter cavity and senses the filter cavity length noise, the filter cavity length noise appears in CC2 loop which controls the relative phase between CCSB and LO. In the case of GW detectors, the LO is the interferometer laser. In this section, we will calculate the CC2 error signal which includes the phase noise from the filter cavity. For simplicity, here we write $\rho_\pm (\Delta\omega_\mathrm{fc}) = \rho_\pm$, $\alpha_\pm (\Delta\omega_\mathrm{fc}) = \alpha_\pm$ and $\alpha_\pm (\Delta\omega_{\mathrm{fc},0}) = \alpha_{\pm,0}$. \par
Signal at OMC transmission is
\begin{eqnarray}
&& P_\mathrm{CC2} \nonumber \\
&=& \left|a_0 e^{i(\omega_0 t + \phi_\mathrm{LO})} + \tau_\mathrm{cc} a_+ \rho_+ e^{i(\omega_0 + \Omega_\mathrm{cc})t
+i(\phi_\mathrm{CC}+\alpha_+)} \right. \nonumber\\
&+&\left. \tau_\mathrm{cc} a_-  \rho_- e^{i(\omega_0 - \Omega_\mathrm{cc})t+i(\phi_\mathrm{CC}+\alpha_- 
+2\delta \phi_\mathrm{pump})}\right|^2 \nonumber \\
&=& 2\tau_\mathrm{cc} a_0a_+ \rho_+ \cos{(\Omega_\mathrm{cc} t -\phi_\mathrm{LO}+\phi_\mathrm{CC}+\alpha_+)} \nonumber \\
&+& 2\tau_\mathrm{cc} a_0a_- \rho_- \cos{(\Omega_\mathrm{cc} t +\phi_\mathrm{LO}-\phi_\mathrm{CC}-\alpha_- 
-2\delta\phi_\mathrm{pump})} \nonumber\\
&+& (\mathrm{DC\ term}) + (\mathrm{2\Omega_\mathrm{cc}\ term}) \label{OMC trans}
\end{eqnarray}
where $a_0$ is the amplitude of LO and $\tau_\mathrm{cc}$ is the transmissivity of CCSB from OPO to OMC transmission. \par
Demodulating (\ref{OMC trans}) by $\cos{(\Omega_\mathrm{cc} t +\theta_\mathrm{dm})}$ and low-passing it,
where demodulation phase $\theta_\mathrm{dm}$ is 
\begin{eqnarray}
\theta_\mathrm{dm} &=& \dfrac{\alpha_{+,0} - \alpha_{-,0}}{2}
\end{eqnarray}
we find
\begin{eqnarray}
P_\mathrm{I} &=& \tau_\mathrm{cc} a_0 a_+ \rho_+ \sin{\left(-\phi_\mathrm{LO}+\phi_\mathrm{CC}+\alpha_{p,0} + \delta \alpha_+ +\frac{\pi}{2}\right)} \nonumber \\
&-& \tau_\mathrm{cc} a_0 a_- \rho_- \mathrm{sin}\left(\phi_\mathrm{LO} - \phi_\mathrm{CC} -  \alpha_{p,0} - \delta \alpha_- - \frac{\pi}{2} \right. \nonumber \\
&-&  \left. 2\delta\phi_\mathrm{pump}\right) \label{CC demod}
\end{eqnarray}
where 
\begin{eqnarray}
\alpha_{p,0} &=&  \dfrac{\alpha_{+,0} + \alpha_{-,0}}{2}\\
 \delta \alpha_{\pm} &=& \alpha_{\pm}  - \alpha_{\pm,0} 
\end{eqnarray}
When the squeezing angle $\phi_\mathrm{sqz}$ is different from $\pi/2$ (squeeze quadrature) by $\delta \phi_\mathrm{sqz}$, (\ref{squeezing angle}) can be written as
\begin{eqnarray}
\phi_\mathrm{sqz} = \phi_\mathrm{LO} - \phi_\mathrm{CC} = \frac{\pi}{2} + \delta \phi_\mathrm{sqz}
\end{eqnarray}
Assuming $\delta \phi_\mathrm{sqz}, \delta \alpha_\pm, \delta \phi_\mathrm{pump} \ll 1$, CC2 error signal (\ref{CC demod}) can be written as
\begin{eqnarray}
P_\mathrm{I}
&=& \tau_\mathrm{cc} a_0a_+ \rho_+ [(1+a\rho)\sin \alpha_{p,0} + \{ -(1+a\rho)\delta\phi_\mathrm{sqz} \nonumber \\
&&+ 2(\delta\alpha_p (\Delta\omega_\mathrm{fc},a,\rho) + a \rho\delta \phi_\mathrm{pump})\}\cos \alpha_{p,0} ] \label{CC LO error}
\end{eqnarray}
where $a = a_-/a_+=x$ is the unbalance of the amplitude of CCSB and $\rho = \rho_-/\rho_+$ is the unbalance of the filter cavity reflectivity of CCSB. $\delta \alpha_p(\Delta\omega_\mathrm{fc},a,\rho)$ is
\begin{eqnarray}
\delta \alpha_p (\Delta\omega_\mathrm{fc},a,\rho) = \dfrac{\delta \alpha_+ + a\rho \delta \alpha_- }{2} \label{CC feedback}
\end{eqnarray}
The first term in (\ref{CC LO error}) is the constant offset, the second term is the relative phase noise between CC and LO which does not include the phase noise coming from the filter cavity (frequency-independent phase noise), the third term is the phase noise of CCSB coming from the filter cavity length noise (frequency-dependent phase noise at the detuning frequency) and the fourth term is the phase noise coming from the relative phase noise of CCSB. The constant offset in (\ref{CC LO error}) should be removed to obtain $\delta \phi_\mathrm{sqz}$.  $\delta \alpha_p (\Delta\omega_\mathrm{fc},a,\rho)$ is the coupling from the CCFC loop which reshapes frequency-dependent phase noise as explained in the following section. 

\subsection{Reshaping of frequency-dependent phase noise}
\label{E}
\begin{figure}[h]
\begin{center}
\includegraphics[width=13cm,bb= 50 70 1170 550]{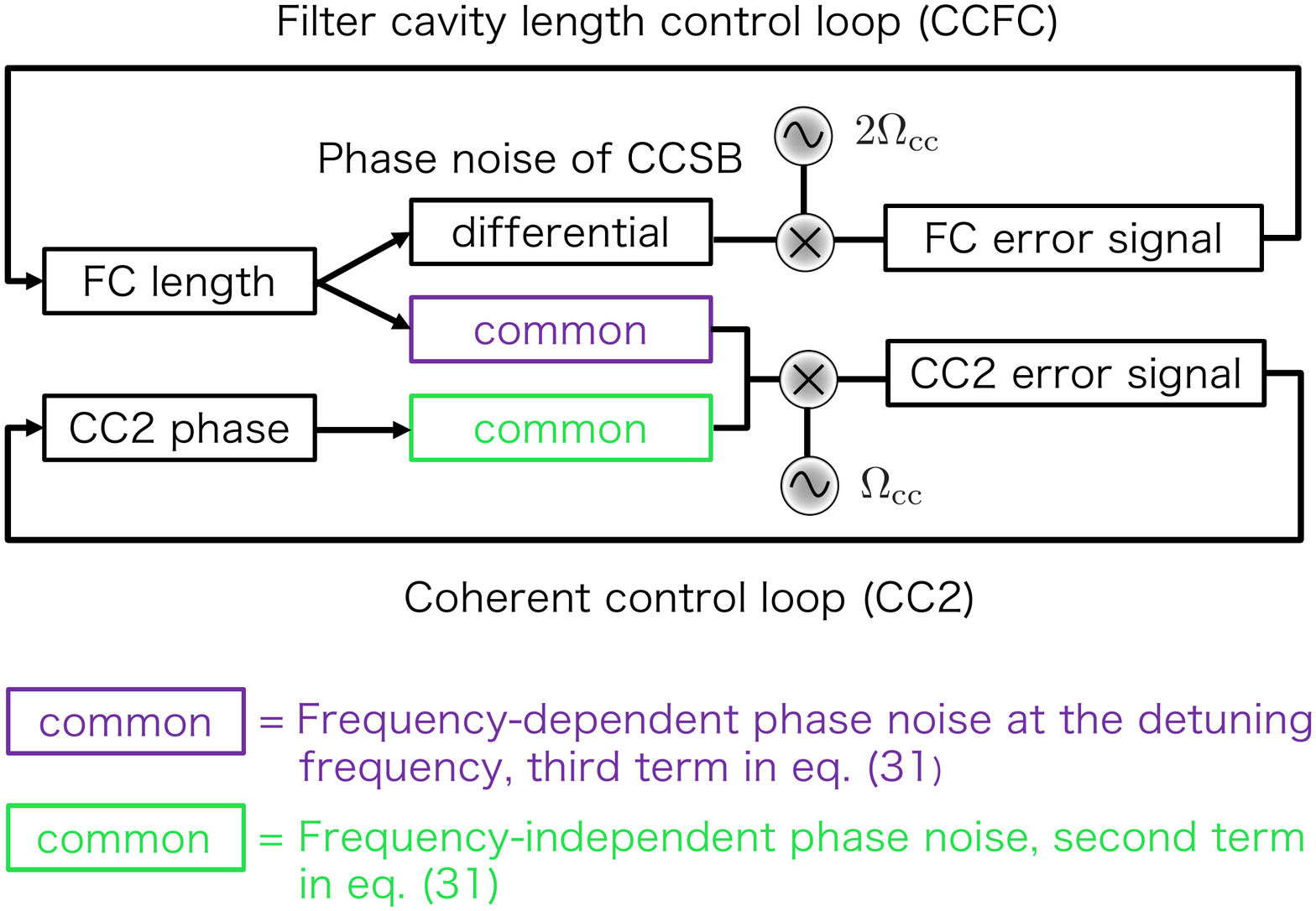}
\caption{Coupling between CCFC loop and CC2 loop.}
\label{coupling}
\end{center}
\end{figure}

The CC2 error signal calculated in Sec. \ref{D} reshapes frequency-dependent phase noise which comes from the filter cavity length noise. This is caused by the coupling between CCFC loop and CC2 loop as shown in FIG. \ref{coupling}. 

The fluctuation of the filter cavity length causes both differential and common phase noise of CCSB. The differential phase noise of CCSB is the CCFC error signal while the common phase noise of CCSB is frequency-dependent phase noise at the detuning frequency, which is the second term in (\ref{CC LO error}). This frequency-dependent phase noise at the detuning frequency couples to the CC2 loop and is suppressed by the feedback loop while the frequency-dependent phase noise is increased at high frequency. In this section, we will explain the detail calculation of the frequency-dependent phase noise with feedback of CC2 loop.
 
Calculation of frequency-dependent phase noise is described in \cite{PhysRevD.90.062006} and the calculation of this section will use the same formalism. For more details of the computation, see Appendix \ref{Appendix B}.\par 
Assuming multiple incoherent noise parameters $X_n$ in quantum noise $\hat{N}$ have small Gaussian-distributed fluctuations with variance $\delta X_n^2$, the average readout noise is given by
\begin{eqnarray}
\hat{N}_\mathrm{avg} \simeq \hat{N} + \sum _n \left(\dfrac{\hat{N}(X_n + \delta X_n)+\hat{N}(X_n - \delta X_n)}{2}-\hat{N}\right)\nonumber\\ \label{phase noise}
\end{eqnarray}
For frequency-independent phase noise, $X_n = \phi$ which is the injection squeezing angle. For the frequency-dependent phase noise, $X_n = \Delta \omega_{\mathrm{fc},0}$ which is the filter cavity detuning.
\par First we will explain about the calculation of frequency-independent phase noise which is necessary in order to calculate frequency-dependent phase noise with feedback from CC2 loop.
The transfer matrix of the squeezer can be written as a function of injection squeezing angle $\phi$,
\begin{eqnarray}
\textbf{S} = \textbf{R}_\mathrm{\phi} \left(
\begin{array}{cc}
e^{\sigma} & 0\\
0 & e^{-\sigma}
\end{array}
\right) \textbf{R}_\mathrm{-\phi}
\end{eqnarray}
where $\textbf{R}$ is a rotation matrix and $e^{-\sigma}$ is injection squeezing level. Frequency-independent phase noise can be represented by variations of injection squeezing angle, $\delta \phi$. 
\par To calculate the effect of phase noise only, we restrict our discussion to an optimally matched filter cavity and no injection, readout losses. Noise due to vacuum fluctuations passing through the squeezer, $\hat{N}_1$ can be written as (see Appendix \ref{Appendix B})
\begin{eqnarray}
\hat{N}_1(\phi) &=& A\cos^2{\phi} + 2B\cos{\phi}\sin{\phi}+ C\sin^2{\phi} \label{N_1 ABC}\\
A &=& (\rho_p^2 e^{-2\sigma}+\rho_m^2 e^{2\sigma})(\cos{\alpha_p}+\mathcal{K}\sin{\alpha_p})^2 \nonumber\\
&+& (\rho_p^2 e^{2\sigma}+\rho_m^2 e^{-2\sigma})(\mathcal{K} \cos{\alpha_p} - \sin{\alpha_p})^2 \\
B &=& (e^{2\sigma}-e^{-2\sigma})(\rho_m^2-\rho_p^2)\nonumber\\
&\times& ( \cos{\alpha_p}+\mathcal{K}\sin{\alpha_p})(\mathcal{K} \cos{\alpha_p} - \sin{\alpha_p})\\
C &=& (\rho_p^2 e^{2\sigma}+\rho_m^2 e^{-2\sigma})(\cos{\alpha_p}+\mathcal{K}\sin{\alpha_p})^2 \nonumber\\
&+& (\rho_p^2 e^{-2\sigma}+\rho_m^2 e^{2\sigma})(\mathcal{K} \cos{\alpha_p} - \sin{\alpha_p})^2
\end{eqnarray}
where 
\begin{eqnarray}
\rho_{^p_m} &=& \dfrac{\rho_+ \pm \rho_-}{2}, \alpha_{^p_m} = \dfrac{\alpha_+ \pm \alpha_-}{2} \nonumber\\
\rho_\pm &=& |r_\mathrm{fc}(\pm \Omega, \Delta\omega_\mathrm{fc})| \\
\alpha_\pm &=& \mathrm{arg}(r_\mathrm{fc}(\pm \Omega, \Delta\omega_\mathrm{fc})) \nonumber
\end{eqnarray}
$\Omega$ is sideband frequency and was fixed to $\Delta \omega_{\mathrm{fc},0}$ in Sections \ref{A}-\ref{D}.
$\mathcal{K}$ is the optomechanical coupling factor of the interferometer,
\begin{eqnarray}
\mathcal{K} = \left(\dfrac{\Omega_\mathrm{SQL}}{\Omega}\right)^2\dfrac{\gamma^2_\mathrm{ifo}}{\Omega^2+\gamma^2_\mathrm{ifo}}
\end{eqnarray}
where $\Omega_\mathrm{SQL}$ is approximately the frequency at which quantum noise reaches the standard quantum limit and $\gamma_\mathrm{ifo}$ is the interferometer bandwidth. In the case of KAGRA, $\Omega_\mathrm{SQL} = 2\pi\times 76.4$ Hz, $\gamma_\mathrm{ifo}= 2\pi\times 382$ Hz \cite{PhysRevD.93.082004}.
When $\phi=0$, $\hat{N}_1(\phi=0) = A$ and this represents the quantum noise of an optimally matched filter cavity with no injection and readout losses, (44) in \cite{PhysRevD.90.062006}.\par
From (\ref{phase noise}), frequency-independent phase noise of $\hat{N}_1$ can be calculated as
\begin{eqnarray}
\hat{N}_{1, \mathrm{avg}} (\phi) = \dfrac{\hat{N}_1(\delta \phi) + \hat{N}_1(-\delta \phi)}{2} = A\cos^2{\delta \phi} + C\sin^2{\delta \phi}\nonumber\\
\end{eqnarray}
Frequency-dependent phase noise can be calculated by averaging $\hat{N}_1(\Delta\omega_{\mathrm{fc},0} + \delta \Delta\omega_\mathrm{fc})$ and $\hat{N}_1(\Delta\omega_{\mathrm{fc},0} - \delta \Delta\omega_\mathrm{fc})$. However, when there is detuning noise, there is also feedback from CC2 loop (\ref{CC feedback}). As shown in FIG. \ref{CC setup}, this feedback from CC2 loop is sent to the squeezer laser and changes the injection squeezing angle $\phi$. Therefore, the frequency-dependent phase noise of $\hat{N}_1$ with feedback from CC2 loop can be calculated as
 \begin{eqnarray}
&&\hat{N}_{1, \mathrm{avg}} (\phi,\Delta \omega_{\mathrm{fc},0})\nonumber\\
&=& \dfrac{1}{2}\left\{ \hat{N}_1(-\delta \alpha_p (\Delta\omega_{\mathrm{fc},0} + \delta \Delta\omega_\mathrm{fc}, a,\rho), \Delta\omega_{\mathrm{fc},0} + \delta \Delta\omega_\mathrm{fc}) \right. \nonumber\\
&&+\left. \hat{N}_1(-\delta \alpha_p (\Delta\omega_{\mathrm{fc},0} -
\delta \Delta\omega_\mathrm{fc},a,\rho),  \Delta\omega_{\mathrm{fc},0}-\delta \Delta\omega_\mathrm{fc})\right\} \nonumber\\ \label{FD phase noise with CC}
\end{eqnarray}
Note that frequency-dependent phase noise $\delta \alpha_p$ is small above cavity pole of the filter cavity $\sim$ 57 Hz and we assumed that the gain of the CC2 loop below 57 Hz is large enough so that the feedback of the CC2 loop is perfect.

\begin{figure*}[t]
\begin{center}
\includegraphics[width=13cm,bb= 00 10 1270 700]{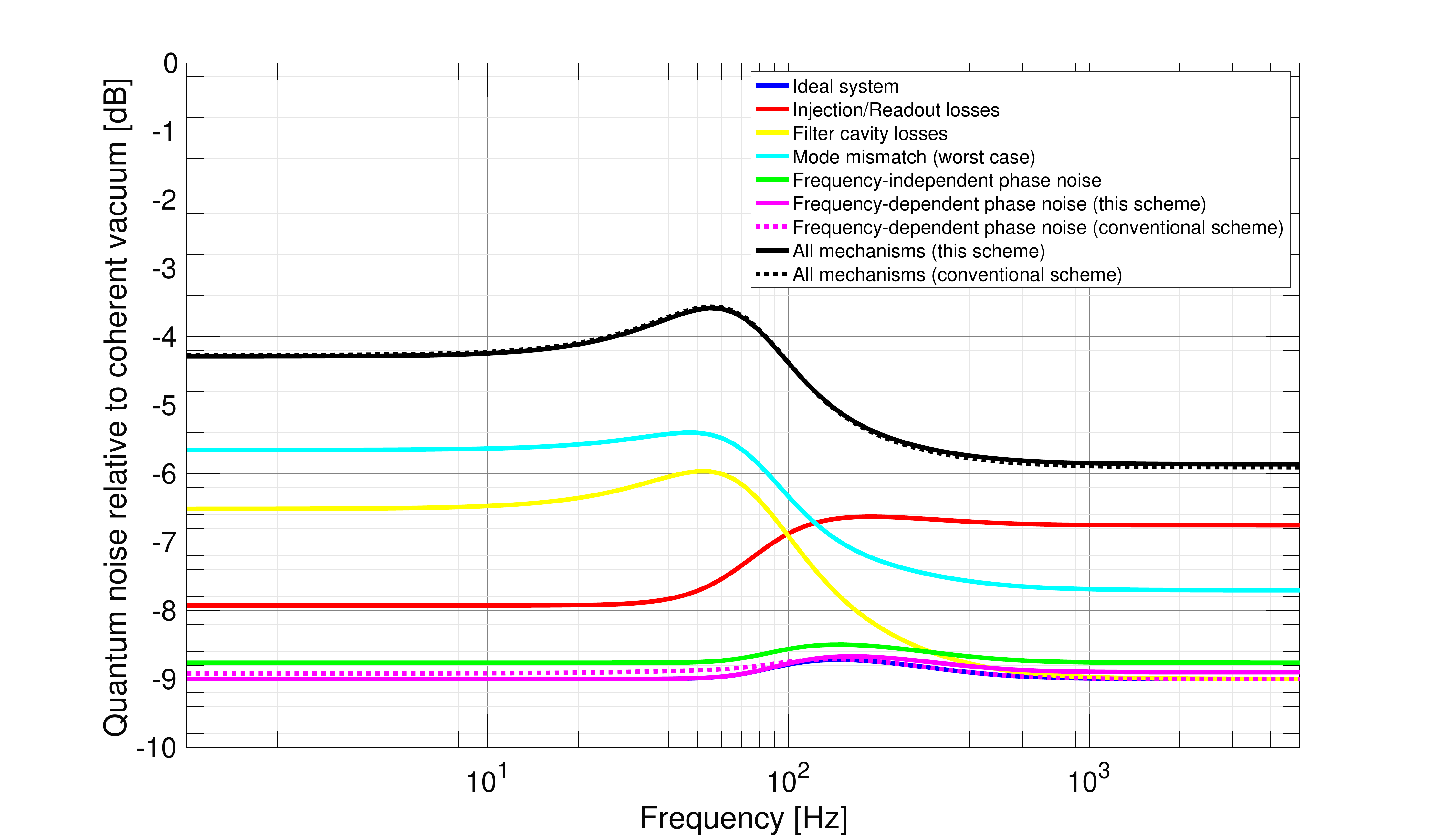}
\caption{Quantum noise relative to coherent vacuum with $\delta L_\mathrm{fc}=1$ pm. Solid purple and black lines are frequency-dependent phase noise and total noise with this scheme. Dotted purple and black lines are frequency-dependent phase noise and total noise with conventional scheme. The solid lines and dotted lines are almost overlapping since the effect of frequency-dependent phase noise is small.}
\label{degradation1}
\end{center}
\end{figure*}

\begin{figure*}[t]
\begin{center}
\includegraphics[width=13cm,bb= 00 10 1270 700]{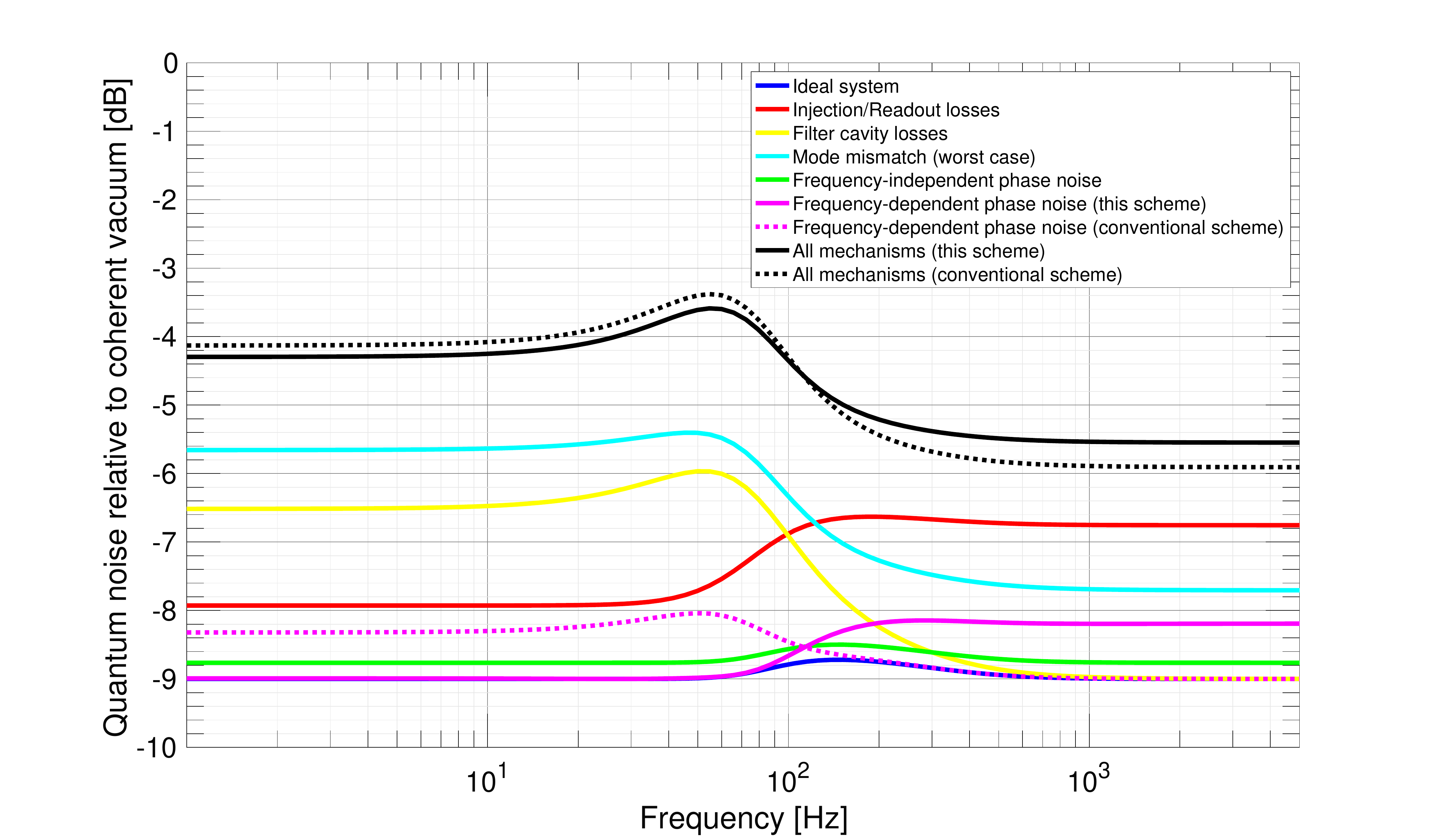}
\caption{Quantum noise relative to coherent vacuum with $\delta L_\mathrm{fc}=3$ pm.}
\label{degradation2}
\end{center}
\end{figure*}

FIG. \ref{degradation1} and \ref{degradation2} show quantum noise relative to coherent vacuum with filter cavity length noise $\delta L_\mathrm{fc}=1$ pm and 3 pm. The frequency-dependent phase noise with this scheme and with conventional scheme is shown as purple line and dotted purple line which are almost overlapping in FIG. \ref{degradation1}. Parameters used in this calculation are shown in TABLE \ref{table1}. The unbalance of the filter cavity relfectivity of CCSB is $\rho = 1.1$ and the unbalance of the amplitude of CCSB is $a=0.47$ ($g=3.6$). As shown in FIG. \ref{degradation1}, frequency-dependent phase noise with this scheme and conventional scheme is small and almost the same with $\delta L_\mathrm{fc}=1$ pm. However, as shown in FIG. \ref{degradation2}, frequency-dependent phase noise with this scheme is suppressed at low frequency by the feedback from CC2 loop while it is increased at high frequency.\par
Effective phase noise at high  frequency with feedback from CC2 loop $\delta \alpha_p$ in (\ref{FD phase noise with CC}) can be calculated from (\ref{phase response}) and (\ref{CC feedback}),
\begin{eqnarray}
&&|\delta \alpha_p (\Delta \omega_{\mathrm{fc},0} \pm \delta \Delta \omega_\mathrm{fc}, a, \rho)| \nonumber \\
&=& \dfrac{\delta \alpha (\Delta \omega_{\mathrm{fc},0})+ a\rho \delta \alpha(-\Delta \omega_{\mathrm{fc},0})}{2} \nonumber \\
&\simeq& 18\ \mathrm{mrad}  \left(\dfrac{1064\ \mathrm{nm}}{\lambda}\right) \left(\dfrac{\mathcal{F}}{4360}\right)\left(\dfrac{\delta L_\mathrm{fc}}{1\ \mathrm{pm}}\right) 
\end{eqnarray}
The squeezing angle is affected by misalignment of LO and CC. The squeezing angle fluctuation at OMC transmission including the misalignment of LO and CC can be written as \cite{Oelker:s}
\begin{eqnarray}
\delta \phi_\mathrm{alignment} \simeq \sum_{ij} A_{ij} \rho^\mathrm{CC}_{ij} \rho^\mathrm{LO}_{ij} \sin{\phi_{ij}}
\end{eqnarray}
where $A_{ij} \sim 1/100$ is the attenuation factor of higher order modes by OMC and $\rho^\mathrm{CC}_{ij}, \rho^\mathrm{LO}_{ij}$ are relative amplitude of CC and LO TEM $ij$ mode with respect to TEM 00 mode. $\phi_{ij} = \phi^\mathrm{LO}_{ij}-\phi^\mathrm{CC}_{ij}$ and $\phi^\mathrm{LO}_{ij}, \phi^\mathrm{CC}_{ij}$ are relative phase of CC and LO TEM $ij$ mode with respect to TEM 00 mode. Considering only HG10/01 mode and assuming that $|\rho^\mathrm{CC}_{ij}|^2 = |\rho^\mathrm{LO}_{ij}|^2 = 10^{-2}$ for $i+j=1$, the squeezing angle fluctuation will be $\delta \phi_\mathrm{alignment} \sim \mathcal{O}(10^{-4})$ rad and it is small enough compared with frequency-independent/dependent phase noise. The misalignment of CCSB also affects the CCFC loop and this effect is calculated in Appendix \ref{Appendix A}.

\section{Noise Calculation}
\label{III}
The requirement of length control of the filter cavity is $\delta L_\mathrm{fc} = 1$ pm and the requirement of alignment control of the filter cavity is $|\gamma|^2 = 0.01$. In this section, we show that shot noise and PLL noise satisfy these requirements. We also show that backscattering noise of leaked carrier from the interferometer to the filter cavity does not spoil the quantum noise above 10 Hz where the quantum noise limits the GW sensitivity.

\subsection{Shot noise}
\subsubsection{Shot noise for length control}
We assume that the power of the coherent control field after OPO is $P_+ = a_+^2 = 1$ uW and the power of the lower coherent control sideband is $P_- = a^2  P_+= 0.22$ uW. When the filter cavity length signal is obtained at OMC reflection, junk light at OMC reflection including higher order modes and other RF sidebands contributes to the shot noise. 
\par Shot noise of CCSB and the junk light at OMC reflection is given by
\begin{equation}
P_\mathrm{shot} = \sqrt{2\hbar\omega_0 (\rho_+^2 P_+ + \rho_-^2 P_- + P_\mathrm{junk})} \simeq  \sqrt{2\hbar\omega_0 P_\mathrm{junk}}
\end{equation}
Here we assumed that frequencies of carrier, CCSB and junk light are almost the same and $\rho_\pm^2 P_\pm \ll P_\mathrm{junk}$. This shot noise within the filter cavity control bandwidth  becomes the filter cavity length noise by the control loop. This shot noise is the most fundamental limit of the filter cavity length signal SNR. From (\ref{FC error rms}), we can compute the maximum power of this junk light at OMC reflection which allows not to spoil the filter cavity length signal as follows,
\begin{eqnarray}
&&\rho_+ \rho_- \sqrt{P_+ P_-}\{\delta \alpha(\Delta \omega_{\mathrm{fc},0}) - \delta \alpha(-\Delta \omega_{\mathrm{fc},0})\} \nonumber \\
&&>  \sqrt{2\hbar\omega_0 P_\mathrm{junk}\Delta f} 
\end{eqnarray}
\begin{eqnarray}
P_\mathrm{junk} &<& 15 \dfrac{P_+ P_-}{\hbar \omega_0\Delta f}\dfrac{\delta L_\mathrm{fc}^2}{\lambda^2} \mathcal{F}^2 \nonumber\\
&=& 15\ \mathrm{W} \nonumber\\
&\times& \left(\dfrac{P_+}{1\ \mathrm{uW}}\right)^2 \left(\dfrac{\mathcal{F}}{4360}\right)^2 \left(\dfrac{\delta L_\mathrm{fc}}{1\ \mathrm{pm}} \right)^2 \left(\dfrac{20\ \mathrm{Hz}}{\Delta f} \right) \nonumber\\ \label{junk light}
\end{eqnarray}
where $\Delta f$ is filter cavity length control bandwidth and set by requirement of backscattering noise \cite{backscattering}. According to (\ref{junk light}), using parameters in TABLE \ref{table1}, $P_\mathrm{junk}<15$ W in order not to spoil the filter cavity length signal and this requirement can be satisfied \cite{PhysRevD.100.082005}.

\subsubsection{Shot noise for alignment control}
We consider only shot noise of displacement signal of the filter cavity and the calculation of tilt signal is entirely analogous. From (\ref{WFS signal}), we can compute the maximum power of junk light at OMC reflection which allows not to spoil the filter cavity alignment signal as follows:
\begin{eqnarray}
\sqrt{\dfrac{2}{\pi}}\sqrt{P_+ P_-} a_\mathrm{WFS} \dfrac{\delta x}{\mathrm{w}_0}>  \sqrt{2\hbar\omega_0 P_\mathrm{junk}\Delta f_\mathrm{WFS}}
\end{eqnarray}
\begin{eqnarray}
P_\mathrm{junk} &<& 0.056 \dfrac{P_+ P_-}{\hbar \omega_0\Delta f_\mathrm{WFS}}\left(\dfrac{\delta x}{\mathrm{w}_0}\right)^2\nonumber\\
&=& 660\ \mathrm{W} \nonumber\\
&\times& \left(\dfrac{P_+}{1\ \mathrm{uW}}\right)^2  \left(\dfrac{(\delta x/\mathrm{w}_0)^2}{0.01}\right) \left(\dfrac{1\ \mathrm{Hz}}{\Delta f_\mathrm{WFS}} \right) \label{WFS junk light}
\end{eqnarray}
where $a_\mathrm{WFS} = 0.42$ is the maximum amplitude of the normalized WFS signal in FIG. \ref{CC WFS} and $\Delta f_\mathrm{WFS}$ is filter cavity alignment control bandwidth. According to (\ref{WFS junk light}), $P_\mathrm{junk}<660$ W in order not to spoil the filter cavity alignment signal and this requirement also can be satisfied.

\subsection{Backscattering noise}
The backscatteting noise comes from the leaked carrier from the interferometer to the filter cavity. The leaked carrier is injected to the filter cavity and scattered by the filter cavity length noise and reinjected into the interferometer. This backscattering noise must be below the vacuum fluctuation in order not to spoil the quantum noise of the interferometer \cite{backscattering}. Considering also the safety factor ($C_\mathrm{safe}=1/10$) and squeezing enhancement factor ($C_\mathrm{sqz} \simeq 1/2$ for 6 dB of quantum noise enhancement), 
\begin{eqnarray}
\delta \alpha (0) \sqrt{P_\mathrm{leak}} &<& C_\mathrm{safe} C_\mathrm{sqz} \sqrt{2\hbar \omega_0} \\
P_\mathrm{leak} &<& 3.1\times10^{-10}\ \mathrm{W} \nonumber \\
&\times& \left(\dfrac{4360}{\mathcal{F}}\right)^2 \left(\dfrac{10^{-16}\ \mathrm{m}/\sqrt{\mathrm{Hz}}}{\delta L_\mathrm{fc}(f)} \right)^2
\end{eqnarray}
where $\delta \alpha (0)$ is phase response of carrier to the filter cavity length noise (\ref{phase response}) and $P_\mathrm{leak}$ is power of the leaked carrier from the interferometer to the filter cavity. We assumed that $\delta L_\mathrm{fc}(f) = 10^{-16}$ m$/\sqrt{\mathrm{Hz}}$ above 10 Hz which can be realized in Advanced LIGO \cite{backscattering}. Given that the carrier output from the interferometer $P_\mathrm{carrier} = 35$ mW in advanced LIGO \cite{backscattering},  81 dB of isolation factor from faraday isolators is required. There is a faraday isolator with more than 40 dB of isolation factor and with less than 1 $\%$ of loss \cite{Genin:18}. By using the 2 faraday isolators, more than 80 dB of isolation factor can be realized with less than 2 $\%$ of loss and the backscattering noise can satisfy the requirement.

\subsection{PLL noise}
The PLL which controls the relative phase between the squeezer laser and the coherent control laser can cause detuning noise. The PLL frequency noise reflected by the filter cavity can be written as
\begin{equation}
S_\mathrm{PLL,fc} (f) = \dfrac{S_\mathrm{PLL} (\phi) f}{\sqrt{1+(f/f_\mathrm{fc})^2}}
\end{equation}
where filter cavity half bandwidth is $f_\mathrm{fc} = 57.3$ Hz and PLL phase noise is $S_\mathrm{PLL} (\phi) = 5\ \mu\mathrm{rad}/\sqrt{\mathrm{Hz}}$ within PLL control bandwidth $\sim$ 40 kHz. The PLL phase noise has been chosen so to have rms of PLL phase noise $\delta \phi_\mathrm{PLL} = 1$ mrad. The rms of PLL frequency noise within the filter cavity control bandwidth is
\begin{eqnarray}
\delta f_{\mathrm{PLL, fc}} &=&\sqrt{\int_{0}^{\Delta f} df {S^2_\mathrm{PLL,fc}(f)}} \nonumber\\
&=& S_\mathrm{PLL}(\phi) \sqrt{\int_{0}^{\Delta f}  df \dfrac{f^2}{1+(f/f_\mathrm{fc})^2}} \nonumber\\
&=& S_\mathrm{PLL}(\phi) f_\mathrm{fc} \sqrt{\left(\Delta f - f_\mathrm{fc} \arctan{\dfrac{\Delta f}{f_\mathrm{fc}}}\right)} \label{PLL rms}
\end{eqnarray}
The rms of PLL frequency noise is $\delta f_{\mathrm{PLL, fc}} = 0.25$ mHz.
This corresponds to rms of the filter cavity length noise $\delta L_{\mathrm{PLL, fc}}  = 2.7\times10^{-16}$ m which satisfies the requirement.\\

\section{Conclusion}
We suggest a new length and alignment control scheme of a filter cavity with coherent control sidebands which are already used to control squeezing angle. It assures accurate detuning and alignment of the filter cavity with respect to squeezed vacuum states. It is shown that coherent control loop reshapes frequency-dependent phase noise with this scheme. The frequency-dependent phase noise at low frequency is suppressed by feedback from the coherent control loop while it is increased at high frequency. We also showed that shot noise and PLL noise with this scheme satisfy the requirement of length and alignment control of the filter cavity and backscattering noise of leaked carrier from the interferometer to the filter cavity does not spoil the quantum noise above 10 Hz where the quantum noise limits the GW sensitivity.

\begin{acknowledgments}
This work was supported by JSPS Grant-in-Aid for Scientific Research (Grants No. 18H01224, No. 18H01235 and No. 18K18763) and JST CREST (Grant No. JPMJCR1873). We thank Kentaro Komori and Keiko Kokeyama for fruitful discussions. 
\end{acknowledgments}

\appendix
\section{Calculation of wavefront sensing}
\label{Appendix A}
Input beam which includes HG 10 mode can be written as
\begin{eqnarray}
E_\mathrm{in} = \left( 
\begin{array}{cc}
U_{00} & U_{10}
\end{array}
\right) \left(
\begin{array}{c}
a_0\\
a_1
\end{array}
\right) E_0 e^{i\omega t}
\end{eqnarray}
where $a_0, a_1$ are the amplitude of HG 00, 10 mode and $U_{00}, U_{10}$ are normalized Hermite-Gaussian modes and can be written as \cite{Kogelnik:66},
\begin{eqnarray}
U_{00}(x,y,z) &=& \sqrt{\dfrac{2}{\pi \mathrm{w}^2(z)}}\mathrm{exp}\left[ -i(kz-\eta (z)) \right. \nonumber\\
&&-  \left. (x^2+y^2)\left(\frac{1}{\mathrm{w}^2(z)}+\frac{ik}{2R(z)}\right) \right]\\
U_{10}(x,y,z) &=& \frac{1}{\sqrt{2}}H_1\left(\dfrac{\sqrt{2}x}{\mathrm{w}(z)}\right)\mathrm{exp}(i\eta(z)) U_{00}
\end{eqnarray}
where $H_1$ is the first order Hermite polynomial, $\mathrm{w}(z)$ is the beam radius, $\eta(z)$ is the gouy phase, $R(z)$ is the beam radius of curvature and can be written as 
\begin{eqnarray}
\mathrm{w}(z) &=& \mathrm{w}_0 \sqrt{1+z^2/z_0^2}\\
\eta(z) &=& \arctan{(z/z_0)}\\
R(z) &=& z(1+z_0^2/z^2)\\
z_0 &=& k\mathrm{w}_0^2/2
\end{eqnarray}
Filter cavity mode which is displaced in $x$ axis by $\delta x$ and tilted around $y$ axis by $\delta\theta$ with respect to input beam can be written as  \cite{Morrison:94}
\begin{eqnarray}
E_\mathrm{fc} &=& \left( 
\begin{array}{cc}
U_{00} & U_{10}
\end{array}
\right)M(\gamma) \left(
\begin{array}{c}
a_0\\
a_1
\end{array}
\right) E_0 e^{i\omega t} \\
M(\gamma)  &=& 
\left(
\begin{array}{cc}
1 & \gamma\\
-\gamma^* & 1
\end{array}
\right)
\end{eqnarray}
where $\gamma$ is (\ref{gamma}) and first order of $\gamma$ is considered. 
\par Reflection matrix of an optimally aligned filter cavity is
\begin{eqnarray}
R^\mathrm{align}_\mathrm{fc}  = 
\left(
\begin{array}{cc}
r_{c0} & 0\\
0 & r_{c1}
\end{array}
\right)
\end{eqnarray}
where $r_{c0},\ r_{c1}$ are, respectively, the filter cavity reflectivity of HG 00 and 10 mode of the coherent control field which is on resonance. Reflection matrix of a misaligned filter cavity in the first order of $\gamma, \gamma_r$ can be written as
\begin{eqnarray}
R^\mathrm{mis}_\mathrm{fc}(\gamma,\gamma_r) &=& M^* (\gamma_r)R^\mathrm{align}_\mathrm{fc} M(\gamma) \nonumber\\
&=& \left(
\begin{array}{cc}
r_{c0} & r_{c0}\gamma + r_{c1}\gamma^*_r\\
 -(r_{c0}\gamma_r + r_{c1}\gamma^*) & r_{c1}
\end{array}
\right)\nonumber\\
\end{eqnarray}
Assuming that $a_0 = 1, a_1 = 0$ for input beam, reflection beam from the misaligned filter cavity can be written as
\begin{eqnarray}
E_\mathrm{ref} = [ U_{00}r_{c0} - U_{10} (r_{c0}\gamma_r + r_{c1}\gamma^*)] E_0 e^{i \omega t} \label{reflection matrix}
\end{eqnarray}
We define $C$ and $S$ as
\begin{eqnarray}
C &=& r_{c0} \gamma_r + r_{c1} \gamma^*  \\
S&=& r_{s0}\gamma_r + r_{s1}\gamma^* \label{S}
\end{eqnarray}
where $r_{s0},\ r_{s1}$ are, respectively, the filter cavity reflectivity of HG 00 and 10 mode of the coherent control sideband which is off resonance.
From (\ref{reflection matrix}) to (\ref{S}), $2\Omega_\mathrm{cc}$ term of the filter cavity error signal (\ref{error signal}) can be written as
\begin{eqnarray}
&& P_\mathrm{cc}(2\Omega_\mathrm{cc}) \nonumber\\
&=& 2a_+ a_-\mathrm{Re} \{ r_+ r_-^* e^{i2\Omega_\mathrm{cc} t}\}\nonumber\\
&=& 2a_+ a_- \mathrm{Re}\{ (U_{00} r_{c0} - U_{10}C)(U_{00}^* r_{s0}^* - U_{10}^*S^*)e^{i2\Omega_\mathrm{cc} t} \}\nonumber \\
&=& 2a_+ a_- \mathrm{Re} \{ (U_{00}U_{00}^*r_{c0}r_{s0}^* - U_{00}U_{10}^*r_{c0} S^* \nonumber\\
&&\hspace{38pt} -\ U_{00}^* U_{10} r_{s0}^*C + U_{10}U_{10}^*C S^*)e^{i2\Omega_\mathrm{cc} t}\} \label{beat WFS}
\end{eqnarray}
\par WFS signal $W$ is given by the sum of the second and third terms of (\ref{beat WFS}). Defining $U = U_{00} U_{10}^\ast$,
\begin{eqnarray}
W &=& 2a_+ a_- \mathrm{Re}(\{ -U r_{c0}(r_{s0}^* \gamma_r^* + r_{s1}^*\gamma)\nonumber\\
&& -\ U^* r_{s0}^* (r_{c0}\gamma_r  + r_{c1}\gamma^*) \}e^{i2\Omega_\mathrm{cc} t}) \nonumber \\
&=&2a_+ a_- \mathrm{Re} (\{-r_{c0} r_{s0}^* (U \gamma_r^* +U^*\gamma_r) \nonumber\\
&&-\ r_{c0}r_{s1}^*U\gamma -r_{c1} r_{s0}^*U^*\gamma^* \}e^{i2\Omega_\mathrm{cc} t})
\end{eqnarray}
Differential signal of $W$ in $x$-axis direction with a quadrant photo diode is 
\begin{eqnarray}
W_\mathrm{diff} = \int \int dx dy \{W(x>0)-W(x<0)\} 
\end{eqnarray}
Since 
\begin{eqnarray}
\int \int dx dy \{U(x>0)-U(x<0)\} = \sqrt{\frac{2}{\pi}} e^{-i\eta(z)}\nonumber\\
\end{eqnarray}
and $r_{c1}, r_{s1} \simeq 1$ due to gouy phase separation in the cavity, WFS signal can be written as (\ref{W QPD}).\par
\begin{figure}[t]
\begin{center}
\includegraphics[width=15cm,bb= 250 0 1670 700]{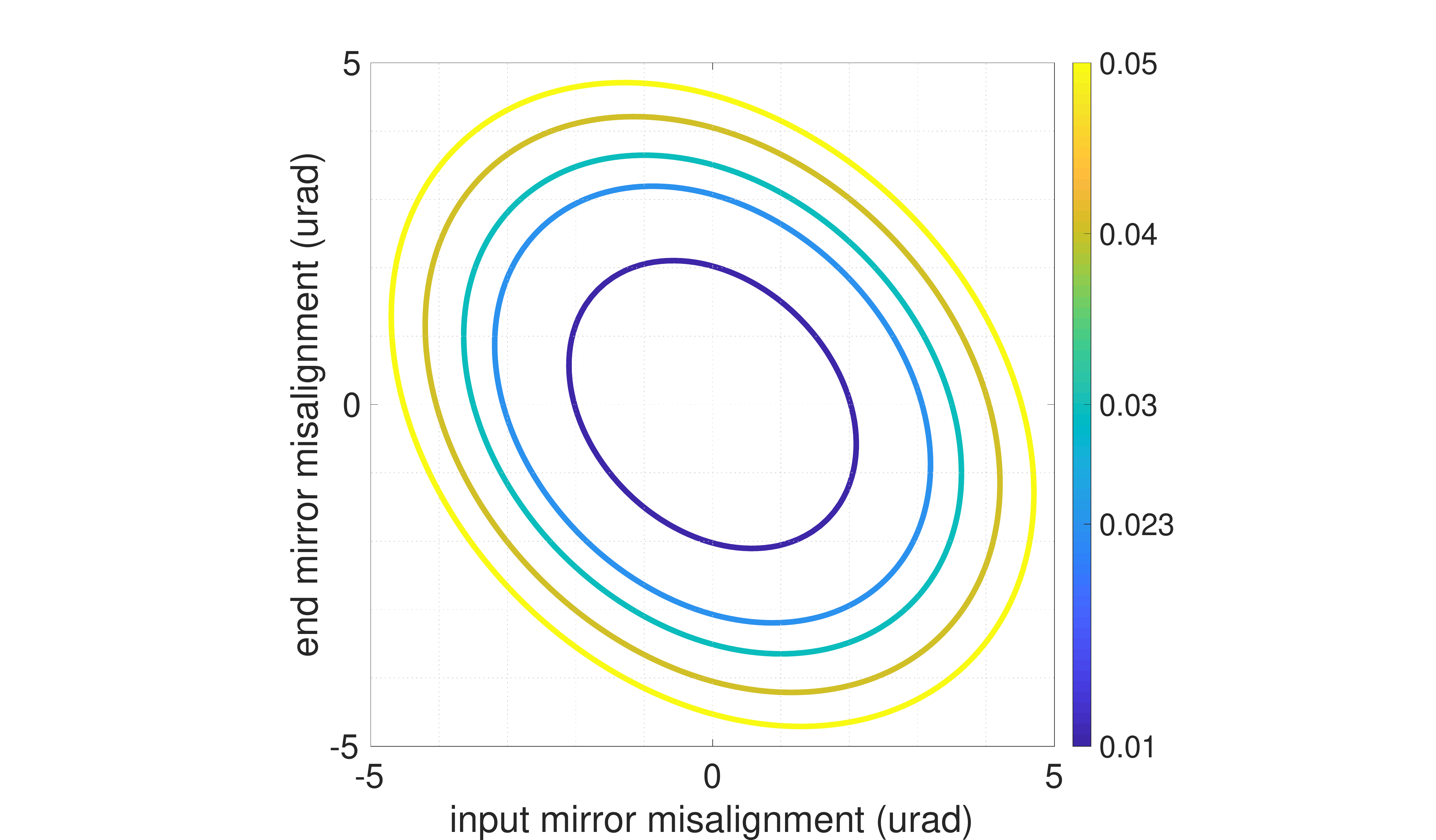}
\caption{Coupling from input, end mirror misalignment of the filter cavity to the filter cavity length signal (\ref{4th term error}) normalized with respect to $a_+a_-$. The coupling should be smaller than the filter cavity length signal $\sim 0.023$. Input, end mirror misalignment should be inside the ellipse corresponding to 0.023.}
\label{4th term contour}
\end{center}
\end{figure}

The fourth term in (\ref{beat WFS}) is a noise source of the filter cavity length signal which is the first term in (\ref{beat WFS}). The fourth term in (\ref{beat WFS}) is
\begin{eqnarray}
&&2a_+ a_- |U_{10}|^2 \mathrm{Re} \{ CS^* e^{i2\Omega_\mathrm{cc}t} \} \nonumber \\
&=& 2a_+ a_- |U_{10}|^2 \mathrm{Re} \{ ( r_{c0}r_{s0}^* |\gamma_r|^2 + r_{c0} r_{s1}^* \gamma_r \gamma + r_{c1} r_{s0}^* \gamma^* \gamma_r^* \nonumber \\
&& + r_{c1}r_{s1}^*|\gamma|^2) e^{i2\Omega_\mathrm{cc}t} \} \label{4th term}
\end{eqnarray}
Demodulating by $\sin{(2\Omega_\mathrm{cc} t -\alpha_{-}(\Delta \omega_{\mathrm{fc},0})})$, the first term in (\ref{4th term}) which is proportional to the filter cavity length signal will disappear. After integration by $x,y$, (\ref{4th term}) will be
\begin{eqnarray}
-a_+ a_- &\{& \mathrm{Re}(r_{c0} \gamma_r \gamma + r_{s0}^*\gamma^*\gamma_r^*)\sin{\alpha_{-}(\Delta \omega_{\mathrm{fc},0})}  \nonumber \\
&+&\mathrm{Im}(r_{c0} \gamma_r \gamma + r_{s0}^*\gamma^*\gamma_r^*)\cos{\alpha_{-}(\Delta \omega_{\mathrm{fc},0})} \nonumber \\
&+& |\gamma|^2\sin{\alpha_{-}(\Delta \omega_{\mathrm{fc},0})} \} \label{4th term error}
\end{eqnarray}
where we used $r_{c1}, r_{s1} \simeq 1$. From (\ref{gamma mirrors}) and (\ref{gamma_r mirrors}), (\ref{4th term error}) can be numerically calculated in terms of input, end mirror misalignment $\delta \theta_I, \delta \theta_E$ as shown in FIG. \ref{4th term contour}. Here we assumed that $R=415$ m and $\mathrm{w}_0 = 0.825$ cm \cite{PhysRevD.98.022010}. (\ref{4th term error}) normalized with $a_+ a_-$ should be smaller than the residual filter cavity length signal normalized with  $a_+ a_-$ which is $P_I/a_+a_- = 23$ mrad. Note that the filter cavity length signal in (\ref{FC error rms}) is normalized with $a_+ a_- \rho_+ \rho_-$. We consider the maximum angular motion of the filter cavity mirrors $\theta_\mathrm{max}$ as $\delta \theta_I^2+\delta \theta_E^2 = \theta_\mathrm{max}^2$ which corresponds to a circle with radius $\theta_\mathrm{max}$ in FIG. \ref{4th term contour}. This circle should be smaller than the ellipse corresponding to 0.023. $\theta_\mathrm{max}$ can be determined from the semi-minor axis of the ellipse corresponding to 0.023 in FIG. \ref{4th term contour}, and we obtain the requirement of $\theta_\mathrm{max}$ as $\theta_\mathrm{max} = 2.7$ $\mu$rad.

\section{Calculation of quantum noise}
\label{Appendix B}
Calculation of quantum noise in this appendix is based on \cite{PhysRevD.90.062006}. Quantum noise can be divided into three parts, noise due to vacuum fluctuations passing through the squeezer $N_1$, noise due to vacuum fluctuations which do not pass through squeezer $N_2$, noise due to vacuum fluctuations in the readout $N_3$.\par
Quantum noise at the interferometer readout is given by
\begin{eqnarray}
N(\zeta) &=& |\overline{\textbf{b}}_\zeta \cdot \textbf{T}_1\cdot v_1|^2+|\overline{\textbf{b}}_\zeta \cdot \textbf{T}_2\cdot v_2|^2+|\overline{\textbf{b}}_\zeta \cdot \textbf{T}_3\cdot v_3|^2\nonumber\\
&\equiv& N_1 + N_2 + N_3 \label{N_1+N_2+N_3}
\end{eqnarray}
where $v_i = \sqrt{2\hbar\omega_0}\textbf{I}$ ($i=1,2,3$) is vacuum fluctuation and $\textbf{I}$ is $2\times2$ identity matrix. $\overline{\textbf{b}}_\zeta = A_\mathrm{LO} (\sin{\zeta}\ \cos{\zeta})$ is local oscillator and  $N(\zeta=0)$ is the quantum noise in the quadrature containing the interferometer signal.\\ $\textbf{T}_1$ is written as
\begin{eqnarray}
\textbf{T}_1 &=& \tau_\mathrm{ro}\textbf{T}_\textbf{ifo}(\textbf{T}_\textbf{00}\textbf{T}_\textbf{fc}+\textbf{T}_\textbf{mm}) \textbf{T}_\textbf{inj} \label{T_1}\\
\textbf{T}_\textbf{ifo} &=& \left(
\begin{array}{cc}
1 & 0\\
-\mathcal{K} & 1
\end{array}
\right) \label{T_ifo}\\
\textbf{T}_\textbf{fc} &=& e^{i\alpha_m}\textbf{R}_{\alpha_p}(\rho_p\textbf{I}-i\rho_m\textbf{R}_{\pi/2})\label{T_fc}\\
\textbf{T}_\textbf{00} &=& |t_\mathrm{00}|\textbf{R}_{\mathrm{arg}(t_\mathrm{00})}, \textbf{T}_\textbf{mm} = |t_\mathrm{mm}|\textbf{R}_{\mathrm{arg}(t_\mathrm{mm})}\\
\textbf{T}_\textbf{inj} &=& 
\tau_\mathrm{inj} \textbf{R}_\mathrm{\phi} \left(
\begin{array}{cc}
e^{\sigma} & 0\\
0 & e^{-\sigma}
\end{array}
\right)\textbf{R}_\mathrm{-\phi} \label{T_inj}
\end{eqnarray}
where $\textbf{T}_\textbf{ifo}$, $\textbf{T}_\textbf{fc}$, $\textbf{T}_\textbf{inj}$ are transfer matrices of interferometer, optimally matched filter cavity, and injection field, respectively.  $\textbf{T}_\textbf{00}$, $\textbf{T}_\textbf{mm}$ are mode matching and mode mismatch matrix. \\
$\tau_\mathrm{inj}$, $\tau_\mathrm{ro}$ is injection, readout transmissivity and can be written as 
\begin{eqnarray}
    \tau_\mathrm{inj} &=& \sqrt{1-\Lambda_\mathrm{inj}^2} \label{tau_inj}\\
    \tau_\mathrm{ro} &=& \sqrt{1-\Lambda_\mathrm{ro}^2}
    \label{tau_ro}
\end{eqnarray}
where $\Lambda_\mathrm{inj}^2$, $\Lambda_\mathrm{ro}^2$ are injection, readout losses.\\
$t_{00}$, $t_\mathrm{mm}$ can be written as $t_{00} = a_0 b^*_0$, $t_\mathrm{mm} = \sum _{n=1} ^{\infty} a_n b^*_n$ where $a_n, b_n$ are complex coefficients when we express the squeezed vacuum states and the local oscillator in the basis of the filter cavity mode and can be written as
\begin{eqnarray}
U_\mathrm{sqz} &=& \sum _{n=0} ^{\infty} a_n U_n, \mathrm{with}\ \sum _{n=0} ^{\infty} |a_n|^2 = 1 \\
U_\mathrm{lo} &=& \sum _{n=0} ^{\infty} b_n U_n, \mathrm{with}\ \sum _{n=0} ^{\infty} |b_n|^2 = 1
\end{eqnarray}
where $U_n$ are the orthogonal basis of spatial modes and $U_0$ is the filter cavity fundamental mode.
\\$\textbf{T}_2$ is written as
\begin{eqnarray}
\textbf{T}_2 &=& \tau_\mathrm{ro}\textbf{T}_\textbf{ifo}\Lambda_2 \\
\Lambda_2 &=& \sqrt{1-(|\tau_2(\Omega)|^2+|\tau_2(-\Omega)|^2)/2} \\
\tau_2 (\Omega) &=& (t_{00}r_\mathrm{fc}(\Omega)+t_\mathrm{mm})\tau_\mathrm{inj}
\end{eqnarray}
$\textbf{T}_3$ is written as
\begin{eqnarray}
\textbf{T}_3 = \Lambda_\mathrm{ro}
\end{eqnarray}
Quantum noise normalized with respect to shot noise level used in Sec. \ref{E} is 
\begin{eqnarray}
\hat{N} = \dfrac{N}{2\hbar \omega_0 A^2_\mathrm{LO}} \label{N hat}
\end{eqnarray}
For calculation of frequency-dependent phase noise, we can consider only $N_1$ since frequency-dependent phase noise of $N_2$ is negligible compared with $N_1$ and $N_3$ is independent of the filter cavity. From (\ref{N_1+N_2+N_3})-(\ref{tau_ro}) and (\ref{N hat}), (\ref{N_1 ABC}) can be derived.\\

\bibliography{CC_PRD}

\end{document}